\documentclass[conference,compsoc]{IEEEtran}
\IEEEoverridecommandlockouts 
\usepackage{xcolor}
\definecolor{hidden-draw}{rgb}{0.8,0.8,0.8}
\definecolor{hidden-pink}{rgb}{1.0,0.7,0.75}
\usepackage{cite}

\ifCLASSINFOpdf
\else
\fi
\usepackage{algorithm}
\usepackage[noend]{algorithmic}
\usepackage[inline]{enumitem}
\usepackage[utf8]{inputenc}
\usepackage{amsmath}
\usepackage{amsfonts}
\usepackage{mathtools}
\usepackage{booktabs} 
\usepackage{xcolor}
\usepackage{comment}
\usepackage{amsmath,graphicx}
\usepackage{multirow}
\usepackage{amssymb}
\usepackage{mathtools,amsthm}
\usepackage{url}
\usepackage{epstopdf}
\usepackage{amsthm}
\usepackage{enumitem}
\usepackage{lipsum}
\usepackage{soul}
\usepackage{subfig}
\usepackage{times}
\usepackage{makecell}
\usepackage{color,soul}
\soulregister\cite7
\soulregister\ref7
\soulregister\pageref7
\usepackage{diagbox}
\usepackage[export]{adjustbox}
\renewcommand{\footnotesize}{\scriptsize}
\usepackage{bm}

\usepackage{soul}
\usepackage{comment}
\usepackage{caption}

\usepackage{lipsum}
\usepackage{microtype}
\newcommand\eat[1]{} 
\usepackage{footnote}
\makesavenoteenv{tabular}
\usepackage{setspace}
\usepackage{titlesec}
\usepackage{graphicx}

\usepackage{multirow}
\usepackage{soul}
\usepackage{xcolor}
\sethlcolor{green!40}
\newcommand{\hlred}[1]{{\sethlcolor{red!20}\hl{#1}}}
\newcommand{\hlyellow}[1]{{\sethlcolor{yellow!40}\hl{#1}}}

\usepackage{caption}
\usepackage{longtable} %

\usepackage{soul}
\usepackage{ulem}
\usepackage[edges]{forest}
\forestset{declare library={forked edges}}

\usepackage{tikz}
\usetikzlibrary{shapes, arrows.meta, positioning}

\tikzstyle{my-box}=[rectangle, draw=hidden-draw, rounded corners, align=left, text opacity=1, minimum height=1.5em, minimum width=5em, inner sep=2pt, fill opacity=.8, line width=0.8pt]
\tikzstyle{leaf-head}=[my-box, draw=gray!80, fill=gray!15, text=black, font=\normalsize, inner xsep=2pt, inner ysep=4pt]
\tikzstyle{red-box}=[my-box, draw=red!70, fill=red!15, text=black, font=\normalsize, inner xsep=2pt, inner ysep=4pt]
\tikzstyle{leaf-red-box}=[my-box, draw=red!80, fill=white, text=black, font=\normalsize, inner xsep=3pt, inner ysep=6pt]

\tikzstyle{blue-box-lv2}=[my-box, draw=cyan!70, fill=cyan!15, text=black, font=\normalsize, inner xsep=2pt, inner ysep=4pt]
\tikzstyle{blue-box-lv3}=[my-box, draw=cyan!70, fill=cyan!15, text=black, font=\normalsize, inner xsep=2pt, inner ysep=4pt]
\tikzstyle{blue-box-lv4}=[my-box, draw=cyan!100, fill=white, text=black, font=\normalsize, inner xsep=3pt, inner ysep=6pt]

\tikzstyle{green-box-lv2}=[my-box, draw=green!70, fill=green!15, text=black, font=\normalsize, inner xsep=2pt, inner ysep=4pt]
\tikzstyle{green-box-lv3}=[my-box, draw=green!70, fill=green!15, text=black, font=\normalsize, inner xsep=2pt, inner ysep=4pt]
\tikzstyle{green-box-lv4}=[my-box, draw=green!100, fill=white, text=black, font=\normalsize, inner xsep=3pt, inner ysep=6pt]

\tikzstyle{orange-box-lv2}=[my-box, draw=orange!70, fill=orange!15, text=black, font=\normalsize, inner xsep=2pt, inner ysep=4pt]
\tikzstyle{orange-box-lv3}=[my-box, draw=orange!70, fill=orange!15, text=black, font=\normalsize, inner xsep=2pt, inner ysep=4pt]
\tikzstyle{orange-box-lv4}=[my-box, draw=orange!100, fill=white, text=black, font=\normalsize, inner xsep=3pt, inner ysep=6pt]

\tikzstyle{violet-box-lv2}=[my-box, draw=violet!70, fill=violet!15, text=black, font=\normalsize, inner xsep=2pt, inner ysep=4pt]
\tikzstyle{violet-box-lv3}=[my-box, draw=violet!70, fill=violet!15, text=black, font=\normalsize, inner xsep=2pt, inner ysep=4pt]
\tikzstyle{violet-box-lv4}=[my-box, draw=violet!100, fill=white, text=black, font=\normalsize, inner xsep=3pt, inner ysep=6pt]

\title{SoK: Are Watermarks in LLMs Ready for Deployment?}

\author{
Kieu Dang\textsuperscript{1},
Phung Lai\textsuperscript{1},
NhatHai Phan\textsuperscript{2},
Yelong Shen\textsuperscript{3},
Ruoming Jin\textsuperscript{4},
Abdallah Khreishah\textsuperscript{2},
My T. Thai\textsuperscript{5}%
\thanks{
\textsuperscript{1}SUNY-Albany, 
\textsuperscript{2}New Jersey Institute of Technology, 
\textsuperscript{3}Microsoft, 
\textsuperscript{4}Kent State University, 
\textsuperscript{5}University of Florida. 
Contact: vdang@albany.edu}
}

\begin{document}
\maketitle

\begin{abstract}
The deployment of a proprietary LLM raises significant intellectual property (IP) violation risks. An adversary can replicate an LLM by using input prompt-the LLM's output pairs to train a surrogate model. It leads to financial setbacks for the LLM's service provider. 
A practical solution for this problem is LLM watermarking, in which the service provider implants an imperceptible pattern (watermark) into the LLM's outputs such that it can detect the watermark from the adversary's generated text without affecting the LLM's utility. While various watermarking techniques have emerged to mitigate these risks, it remains unclear how far the community and industry have progressed in  developing and deploying watermarks   in LLMs.


 
To bridge this gap, we aim to develop a comprehensive systematization for WMs in LLMs by 1) presenting a detailed taxonomy for WMs in LLMs, 2) proposing a novel IP classifier to explore the effectiveness and impacts of WMs on LLMs under both attack and attack-free environments, 3) analyzing the limitations of existing WMs in LLMs, and 4) discussing practical challenges and potential future directions for WMs in LLMs. Through extensive experiments, we show that despite promising research outcomes and significant attention from leading companies and community to deploy WMs, these techniques have yet to reach their full potential in real-world applications due to their unfavorable impacts on model utility of LLMs and downstream tasks. Our findings provide an insightful understanding of WMs in LLMs, highlighting the need for practical WM solutions tailored to LLM deployment.

\end{abstract}


%
\IEEEpeerreviewmaketitle

\section{Introduction}

Large language models (LLMs), such as ChatGPT, Gemini, Claude,  and Cohere \cite{gemini,OpenAI,AnthropicClaude,Cohere}, have demonstrated remarkable capabilities in text generation, machine translation, and knowledge understanding tasks \cite{zhang2023prompting,zhang2023machine,xu2023paradigm,hu2023unbiased,kirchenbauer2023watermark}. They effectively mimic human writing behaviors and generate complex and coherent outputs from the input text, making it challenging to determine whether a text is authored by humans or generated by LLMs. Due to the high demands of computational resources and human efforts required for training LLMs \cite{brown2020language,chen2021evaluating}, these models are commonly offered as a service through  application programming interfaces (APIs), typically requiring users to pay or subscribe \cite{minthigpen,ibm}. Although users  cannot access to the model weights or architectures of these commercial LLMs, this restriction does not ensure  the safety of these models. Malicious actors can intentionally mimic cloud-hosted LLM behaviors to offer cheaper services \cite{wallace2020imitation,xu2021student}. To  conduct this  service stealing, an adversary can query a set of inputs through an LLM's API to retrieve the corresponding outputs. Then, the adversary uses these input-output data to  effectively fine-tune their local model. When the number of queries is sufficient  to gather enough input-output data within a particular domain, the adversary can steal the cloud-hosted LLM behaviors  in that domain \cite{googlebard,carlini2024stealing,aiattack}.  Consequently, these concerns underscore significant risks regarding the intellectual property (IP) rights of the cloud-hosted proprietary LLMs \cite{li2023protecting}. 

To address such risks, service providers, i.e., the cloud, have employed many strategies including watermarks, encryption, limited model exposure via  APIs, and differential privacy \cite{kirchenbauer2023watermark,xue2022advparams,yu2021differentially}. Among them,  watermarks  (WMs) \cite{kirchenbauer2023watermark,yoo2023advancing,liu2023unforgeable,christ2024undetectable,kuditipudi2023robust} have emerged as a practical tool for LLMs  due to its ability to ensure traceability, protect  IP, and detect misuse. Typically, WMs embed imperceptible patterns   into the LLM outputs directly. These patterns can be used to trace the original text or determine whether the generated text from an LLM is watermarked, facilitating the detection of unauthorized use. To watermark outputs, the cloud can introduce bias into logits of token generations to favor a set of specific tokens or alter the token sampling process of LLMs.
Once an WM is applied, the cloud can use IP checkers to determine if a set of outputs from a suspicious model is watermarked. This allows the cloud to assess whether 
the suspicious model has been trained or fine-tuned using the cloud's watermarked outputs. Typically, IP checkers analyze the tokens and perform statistical tests to check if the values exceed certain thresholds, indicating that the cloud-hosted LLM has been imitated.

WMs are effective in detecting IP violations. However, 
WMs can be susceptible to WM removal attacks \cite{zhang2023watermarks, pang2024attacking} or spoofing attacks \cite{pang2024attacking}. These attacks either require significant computational resources for additional training to paraphrase the generated text of cloud-hosted LLMs or demand numerous queries to learn the distribution of WMs.  While several studies have surveyed WMs on LLMs \cite{liu2024survey,liang2024watermarking,lalai2024intentions}, no research has systematically and extensively explored the effectiveness of WMs on LLMs, particularly regarding the impact of WMs on LLM utility and IP checkers, under both attack and attack-free environments. In addition, the practicality of WMs has not been rigorously assessed. Therefore, in this work, we aim to develop a comprehensive systematization of WMs in LLMs, with the following contributions:

\begin{itemize}
    \item We present a taxonomy for WMs that effectively mitigate model stealing attacks.
\item We propose a novel cross-model  IP classifier to explore the effectiveness and impacts of WMs on LLMs, given both attack and attack-free environments.
\item We suggest  well-suited applications and limitations of different WMs in LLMs.
\item We conduct extensive experiments to demonstrate the effectiveness and practicability of WMs in LLMs.
\item We discuss limitations, challenges, barriers, and future directions for WMs in LLMs.  
\end{itemize} 
 Our study provides key insights into using WMs for protecting LLMs, addressing concerns related to their deployment and IP protection.
 Through extensive experiments across different WMs, LLMs, and IP checkers, our highlighted findings are: \textbf{(1)} WMs significantly enhance the uniqueness and distinctiveness of LLM outputs, showing their effectiveness in strengthening IP protection and reducing unauthorized usage; \textbf{(2)}  The impact of WMs on model utility can be moderate to significant, varying based on WM types and LLM architectures, potentially influencing their suitability for real-world applications; and \textbf{(3)} Attacks targeting WMs impose a notable cost on model utility. In conclusion, due to their unfavorable impacts on model utility, WMs in LLMs are not ready for real-world deployments, highlighting the urgent need to improve WM resilience while maintaining model utility.

\section{Large Language Models and  Risks} 
\label{sec: LLM and risks}

\subsection{Large Language Models}

Large language models (LLMs)   \cite{radford2019language,OpenAI, MicrosoftAzureAI, HuggingFaceMetaLLaMA} are advanced deep learning models, pre-trained on vast datasets with  a large number  of parameters, enabling them to perform a wide range of natural language processing (NLP) tasks, such as text classification, text and code generation, and question answering. 

Given a vocabulary $\mathcal{V}$ consisting of words or word fragments, referred to as tokens, a sequence  $x = \{x^{(i)}\}_{i=1}^T \in \mathcal{V}^T$ represents a \textit{prompt} of   $T$  tokens, and $y =  \{y^{(i)}\}_{i=T+1}^K$ denotes the tokens generated by an LLM in response to the prompt $x$ ($T$ and  $K$ are natural numbers). An LLM model $\theta$ is trained to maximize the probability of the  token sequence $y$, conditioned on the prompt $x$, as follows:
\begin{equation}
    P(y|x)= \Pi_{i=1}^T P(y^{(i)}|x^{(1)},x^{(2)}, \cdots, x^{(i-1)})
    \label{eq:llm}
\end{equation}
In generation tasks, the model  predicts each token  in an autoregressive manner \cite{radford2019language}, similar to Eq.~\ref{eq:llm}, to generate a complete text sequence. 

\begin{figure}[t]
      \centering    
      \includegraphics[scale=0.125]{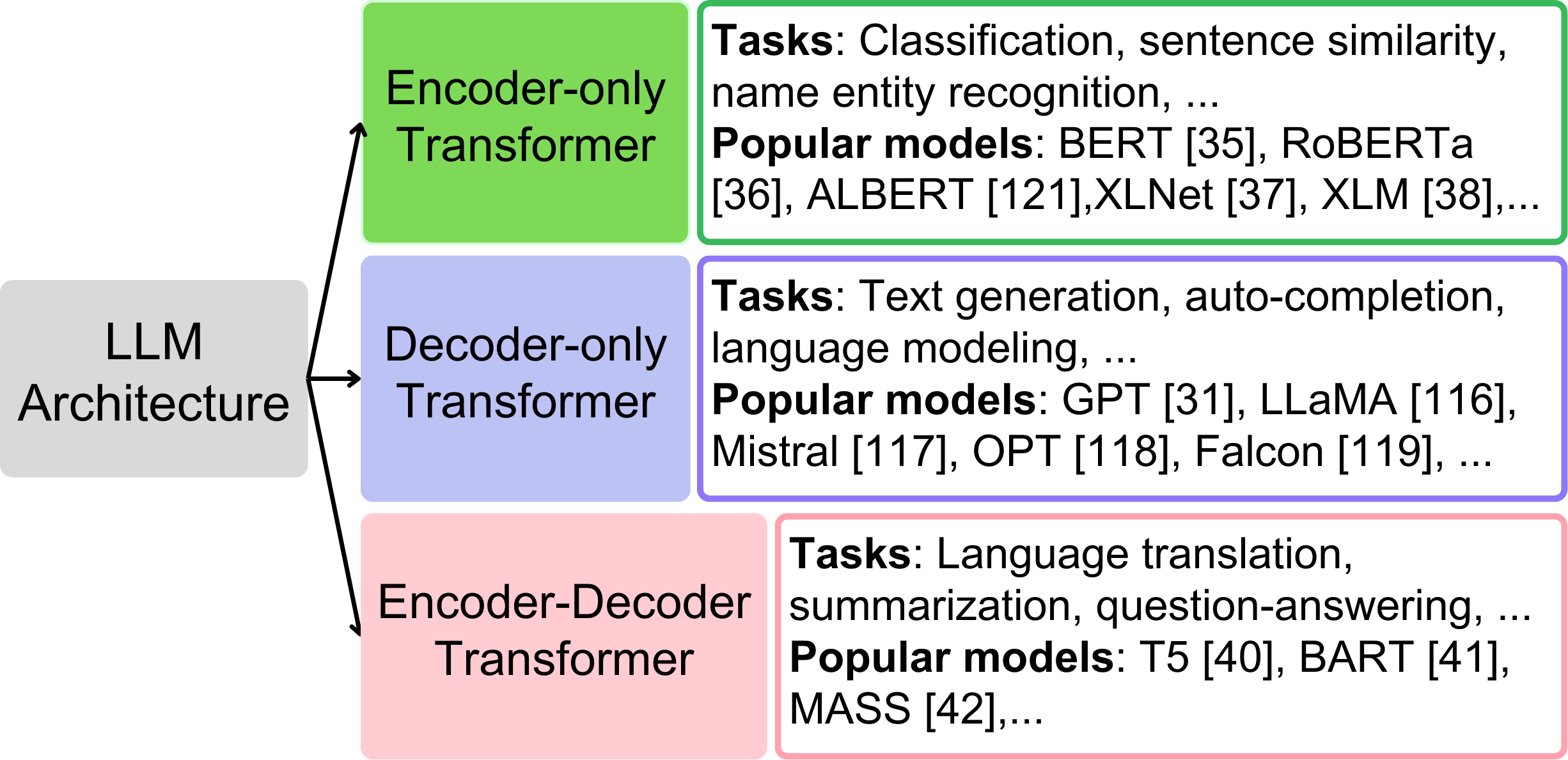} 
      \caption{Categories of Large Language Models.} 
      \label{fig:llm_architecture}
 \end{figure}

The most widely used LLM architectures are based on transformers \cite{vaswani2017attention} and incorporate a self-attention mechanism. This mechanism includes a stack of multi-head attention and feed-forward layers. Positional encoding with distinct matrices for different positions ensures the model retains sequence information without relying on recurrent networks. 
Based on transformer architectures, existing LLMs can be divided into three main categories (Fig.~\ref{fig:llm_architecture}), as follows. 
(1) \textit{Encoder-only models}, such as BERT \cite{Devlin2019BERTPO}, RoBERTa \cite{liu2019roberta}, XLNet \cite{yang2019xlnet}, or XLM \cite{conneau2019cross}, capture word context and are well-suited for tasks that require a deep understanding of input data, such as  sentence classification, sentiment analysis, and information extraction. These models are trained bidirectionally, predicting masked words using both preceding and following words. 
 \textit{(2) Decoder-only models}, such as GPT and its family \cite{Radford2018ImprovingLU}, operate unidirectionally, predicting the next token based on  context in an auto-regressive manner. They are designed for text generation tasks  without a separate encoding phase. 
Lastly, (3) \textit{Encoder-decoder models} such as T5 \cite{raffel2020exploring}, BART \cite{DBLP:journals/corr/abs-1910-13461} or MASS \cite{song2019mass}
combine both encoder and decoder components for tasks that involve understanding and generating data, such as text summarization and translation. 


 \vspace{-1em}

\subsection{Risks in LLMs}

Due to the high demands of computational resources and human efforts required for training LLMs \cite{brown2020language,chen2021evaluating}, LLMs are commonly offered as a service through APIs, typically requiring users to pay or subscribe \cite{minthigpen,ibm}.
 Table \ref{table:llmAPI2} (Appendix) presents a comprehensive overview of popular LLM APIs, highlighting their  features, advantages, and constraints. According to the Artificial Analysis leaderboard \cite{leaderboard}, 
 OpenAI and Microsoft are currently the leading providers in the LLM API market.

 Although users may not have access to the model weights or architectures via APIs, this restriction does not guarantee the model's safety. In addition, as LLMs become more widespread, the risk of their misuse for malicious purposes is increasing. Potential misuses of LLMs include several significant threats, as follows.
First, LLMs can be exploited to generate false information or carry out harmful actions, such as crafting phishing emails, hate speech, or biased content to manipulate public opinion or amplify harmful narratives \cite{chen2023can}. Second, LLMs can generate deepfake data, making manipulated media more convincing and enabling sophisticated scams or damage to reputation \cite{mitra2024world}. Third, adversaries can use LLMs to mimic individuals or organizations in text-based communications, facilitating theft, fraud, or financial crimes \cite{chen2023combating}. Fourth, LLMs can assist in writing malicious code or crafting evasion techniques to bypass detection systems, posing a significant cybersecurity  threat \cite{motlagh2024large}. More importantly, malicious actors may intentionally exploit LLM behaviors by imitating cloud-hosted models to offer cheaper services or engage in model-stealing attacks, further amplifying the aforementioned risks. Such attacks not only threaten the security of proprietary models but also expose users to fraudulent or unethical services \cite{wallace2020imitation,xu2021student}.




%
   

 \begin{figure}[t]
      \centering    
      \includegraphics[scale=0.13]{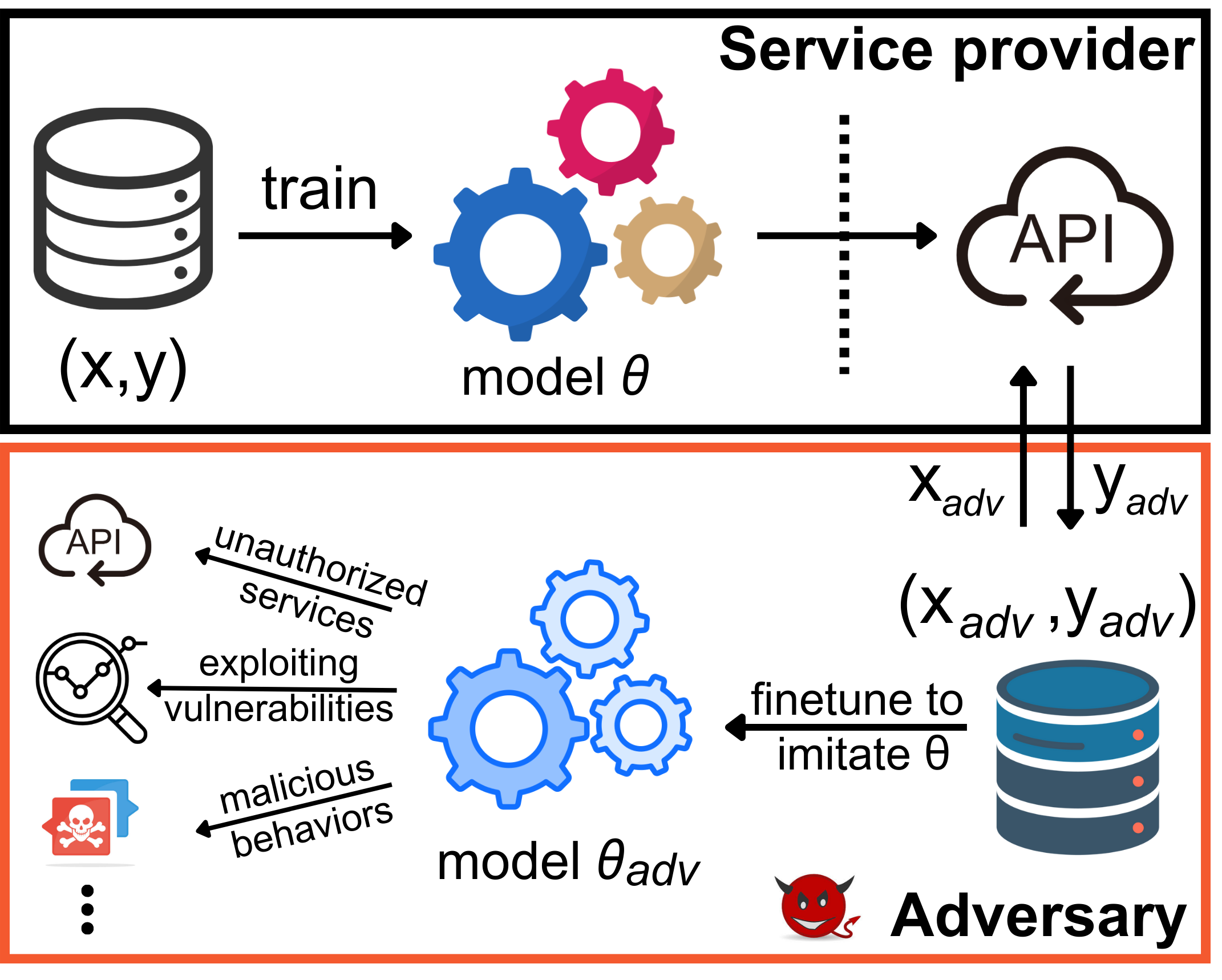} 
      \caption{Model Stealing Attacks \cite{jovanovic2024watermark,dang2025delta}.} 
      \label{fig:stealingattack}
 \end{figure}

\subsection{Model Stealing Attacks }

In model stealing attacks \cite{jovanovic2024watermark,dang2025delta}, adversaries aim to imitate the behavior of a  cloud-hosted model denoted as  $\theta$, by constructing their local model $\theta_{adv}$ through fine-tuning on input-output pairs obtained by querying the cloud-hosted model \cite{zhang2023apmsa}.  
The goal is to bypass IP checkers while maintaining the adversary model's utility, which is comparable to that of the original model. To launch the attacks (Fig.~\ref{fig:stealingattack}), adversaries start with a set of $N$ task-specific prompts $\{x_i\}_{i=1}^N$, sending them to the cloud to obtain outputs $\{y_i\}_{i=1}^N$. These pairs are used to fine-tune $\theta_{adv}$ to mimic the 
behaviors of $\theta$. 


Model stealing attacks have various malicious and unethical uses, especially in LLMs. Common usages include 1) imitating proprietary models  to resell or compete with the original cloud, undercutting legitimate providers and offering cheaper but unlicensed services, 2) bypassing service fees, 3) reversing engineering and exploiting vulnerability, as stolen models can be analyzed to identify weaknesses and potential vulnerabilities, and then exploit them, 4) violating privacy by inferring if sensitive  data was used in training, 5) enabling malicious behaviors, such as spreading disinformation, biased content, harmful outputs, and 6) compromising IP checkers by copying the IP embedded in the model's architecture, training data, and fine-tuning processes. These attacks pose a significant threat to the security, revenue, and ethical deployment of LLMs, underscoring the need for developers to implement robust defenses mitigating these risks.

\begin{figure*}[t!]
\centering
\resizebox{\textwidth}{!}{%
\begin{forest}
forked edges,
for tree={
    grow=east,
    reversed=true,
    anchor=base west,
    parent anchor=east,
    child anchor=west,
    base=center,
    rectangle,
    rounded corners,
    align=left,
    text centered,
    minimum width=4em,
    edge+={gray, line width=1pt},
    s sep=3pt,
    inner xsep=2pt,
    inner ysep=3pt,
    line width=0.8pt,
},
where level=1{text width=10em,font=\normalsize,}{},
where level=2{text width=10em,font=\normalsize,}{},
where level=3{text width=11em,font=\normalsize,}{},
where level=4{text width=7em,font=\normalsize,}{},
[\textbf{Watermarking} \\ \textbf{for LLMs}, leaf-head
  [\textbf{WM During} \\ \textbf{Logits Generation}, blue-box-lv2
    [{\textbf{Mechanism:} KGW \cite{kirchenbauer2023watermark}, SIR \cite{liu2024semanticinvariantrobustwatermark}, DiPmark \cite{wu2024dipmark}, SemaMark \cite{ren-etal-2024-robust}, Adaptive WM \cite{liu2024adaptive}, \\ Unbiased WM \cite{hu2024unbiased}, TSW \cite{pmlr-v235-huo24a}, WatME \cite{liang2024watme}, GumbelSoft \cite{fu2024gumbelsoft}, MPAC \cite{yoo2024advancing}, UPV \cite{liu2023unforgeable}, Unigram \cite{zhao2024provable}, \\ Topic-based WM \cite{nemecek2024topic}, REMARK LLM \cite{zhang2024remark}, SWEET \cite{lee2023wrote}, CTWL \cite{wangtowards}, EWD \cite{lu2024entropy}, X-SIR \cite{he2024can}, \\ SW \cite{fu2024watermarking}, CodeIp \cite{guan2024codeip}, OW \cite{wouters2023optimizing}, Stylometric WM \cite{niess2024stylometric}, NS-WM \cite{takezawa2023necessary}.\\
    \textbf{Strengths:} Flexibility and non-intrusive WM, Effectively track watermarked text.\\
    \textbf{Weaknesses:} Possible impact on semantic meaning, Vulnerable to removal attacks.
    }, blue-box-lv4, text width=42.5em]
  ]
  [\textbf{WM During} \\ \textbf{Sampling Process}, green-box-lv2
    [\textbf{Token-based} \\ \textbf{Sampling}, green-box-lv3
      [{\textbf{Mechanism:} Undetectable WM \cite{christ2024undetectable}, EXP \cite{kuditipudi2023robust}, EDRW \cite{golowich2024edit},\\ EXPGumbel \cite{aaronson2022watermarking}, PDW \cite{fairoze2023publicly}, Bileve \cite{zhou2024bileve}, STA-1 \cite{mao2024watermark}, SynthID \cite{dathathri2024scalable}.\\
       \textbf{Strengths:} Can be incorporated easily and less noticeably,  Simple\\ detection, checking the alignment between tokens and the sampling pattern.\\
      \textbf{Weaknesses:}   Can be vulnerable to simple text modifications,  \\Negative impact of randomness to generated output. }, green-box-lv4, text width=31em]
    ]
    [\textbf{Sentence-based} \\ \textbf{Sampling}, green-box-lv3
      [{\textbf{Mechanism:} SemStamp \cite{hou2023semstamp}, k-SemStamp \cite{hou2024k}.\\
      \textbf{Strengths:} More resilient to small edits.\\
      \textbf{Weaknesses:} Require more phases, resources and complex algorithms to \\train;   Limited scope: Works only when appropriately trained;   \\ Susceptible to sentence-reordering attacks.\\}, green-box-lv4, text width=31em]
    ]
  ]
  [\textbf{WM During} \\ \textbf{LLM Training}, orange-box-lv2
    [\textbf{Trigger-based} \\ \textbf{Watermarking}, orange-box-lv3
      [{\textbf{Mechanism:} CodeMark \cite{sun2023codemark}, CoProtector \cite{sun2022coprotector}, Hufu \cite{xu2024hufu}, WLM \cite{guwatermarking}, \\ PLMmark \cite{li2023plmmark}.\\
      \textbf{Strengths:} Controlled via trigger, Custom response.\\
      \textbf{Weaknesses:} Ineffective without trigger, Potential misuse if discovered.}, orange-box-lv4, text width=31em]
    ]
    [\textbf{Global} \\ \textbf{Watermarking}, orange-box-lv3
      [{\textbf{Mechanism:} Distillation WM \cite{gu2023learnability}, RLWM \cite{xu2024learning}, EmMark \cite{zhang2024emmark}.\\
      \textbf{Strengths:} Universal and integrated protection.\\
      \textbf{Weaknesses:} Complicated to train, May affect model utility.}, orange-box-lv4, text width=31em]
    ]
    [\textbf{LLM Architecture} \\ \textbf{Watermarking}, orange-box-lv3
      [{\textbf{Mechanism:} Cross-Attention WM \cite{baldassini2024cross}.\\
      \textbf{Strengths:} Robust to synonym substitution and paraphrasing.\\
      \textbf{Weaknesses:} Challenge in balancing WM robustness with the \\ quality of the generated text,  Require fine-tuning, adding complexity to \\ the model’s deployment.\\}, orange-box-lv4, text width=31em]
    ]
  ]
  [\textbf{Miscellaneous}, violet-box-lv2
    [\textbf{Mixed Method} \\ \textbf{Watermarking}, violet-box-lv3
      [{\textbf{Mechanism:} Duwak \cite{zhu2024duwak}, Waterpool \cite{huang2024waterpool}.\\
      \textbf{Strengths:} Higher robustness, Improved detection.\\
      \textbf{Weaknesses:} Increased computational overheads, Potential \\interference between methods.}, violet-box-lv4, text width=31em]
    ]
    [\textbf{Multiple-Output} \\ \textbf{Watermarking}, violet-box-lv3
      [{\textbf{Mechanism:} WaterMax \cite{giboulot2024watermax}.\\
      \textbf{Strengths:} Low quality impact.\\
      \textbf{Weaknesses:} Slower generation, More compute.}, violet-box-lv4, text width=31em]
    ]
    [\textbf{Prompt-Based} \\ \textbf{Watermarking}, violet-box-lv3
      [{\textbf{Mechanism:} ModelShield \cite{pang2024adaptive}.\\
      \textbf{Strengths:} Adaptive, High text fidelity.\\
      \textbf{Weaknesses:} Increased computational overheads,  Vulnerable to \\ WM removal attack.}, violet-box-lv4, text width=31em]
    ]
  ]
]
\end{forest}
}
\caption{Taxonomy of Watermarking Mechanisms for LLMs.}
\label{figure:WMtaxonomy}
\end{figure*}
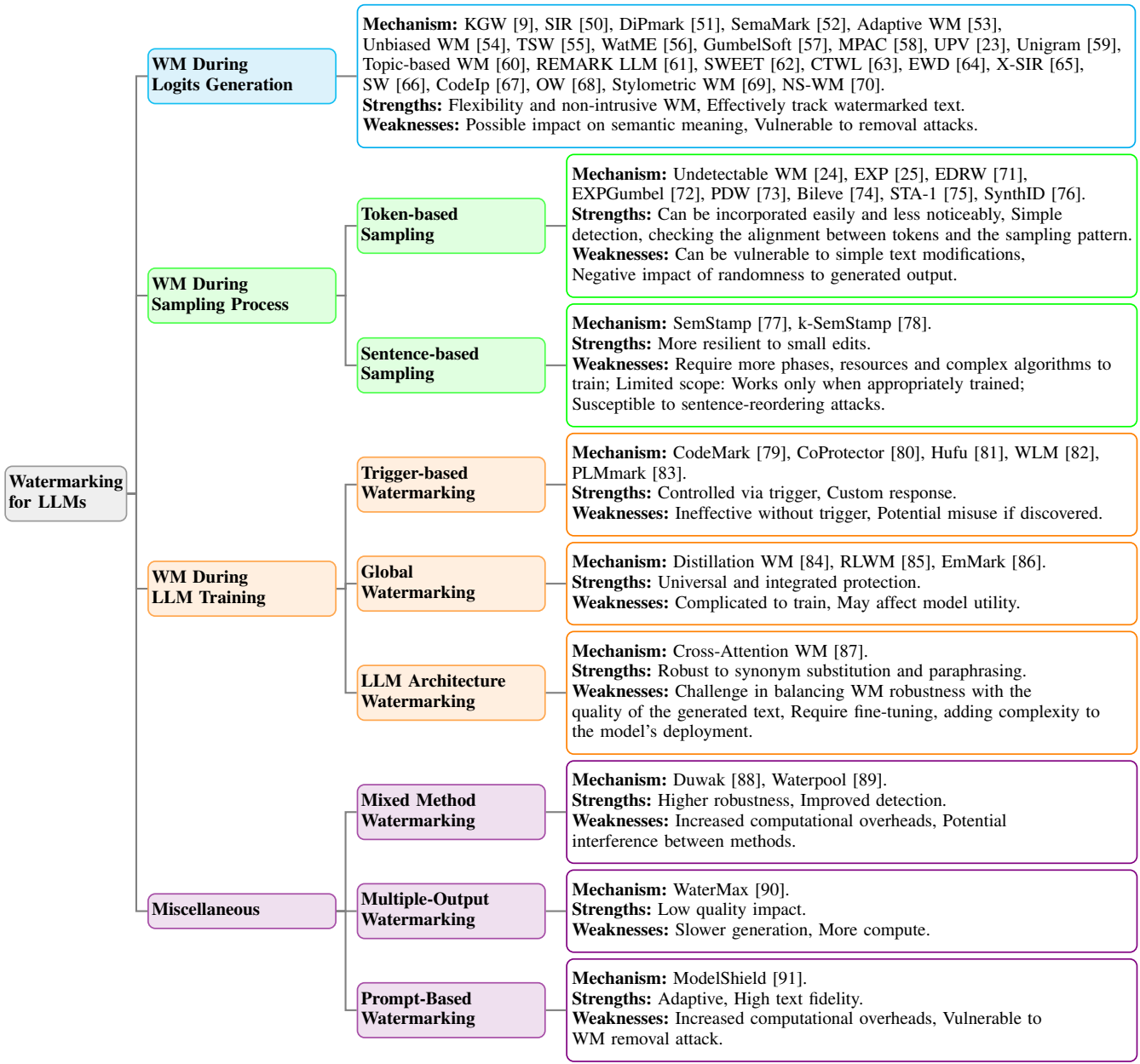

\section{Taxonomy of WMs and IP Checkers} 
\label{sec:WMtaxonomy}


To defend against model stealing attacks, recent works have highlighted the effectiveness and feasibility using WMs   \cite{kirchenbauer2023watermark, kuditipudi2023robust, liu2024semanticinvariantrobustwatermark}. 
 WMs embed a unique and detectable pattern into LLM outputs to trace and identify the origin of the text. 
A WM can be represented as a triple  $(\mathcal{G}, \mathcal{W}, \mathcal{D})$, where $\mathcal{G}$ is the WM generator, which takes a  prompt $x$ and a WM function $\mathcal{W}$ to produce a watermarked output $y^{wm} =  \mathcal{G}(x, \mathcal{W})$, and  $\mathcal{D}$ is a WM detection function, also known as an IP checker, to determine whether a given text is watermarked or not. 
 




\subsection{WM Generators}

WM generators typically introduce perturbations to the probability distribution of LLMs during output generation. Depending on $\mathcal{W}$ and types of perturbations, WMs can be  classified into several research directions, as shown in Fig.~\ref{figure:WMtaxonomy}. \textit{First}, WM during logits generation approaches  partition tokens into green and red lists and add bias into logits to favor tokens from the green list. The tokens in the watermarked output are thus sampled with higher probability from the green list, which is determined by a pseudo-random generator seeded by the input prompt \cite{kirchenbauer2023watermark}. 
These methods are flexible and effective for tracking watermarked outputs but may impact the semantic meaning of the output and are vulnerable to removal attacks  \cite{zhang2023watermarks}.
\textit{Second}, WM during sampling process methods alter LLM text generation's sampling process of next tokens or whole sentences without altering the underlying distribution of output.
\cite{kuditipudi2023robust, hou2023semstamp}. While they are easy to integrate and allow for simple detection, they can be susceptible to attacks such as text modifications, randomness, and sentence reordering, especially in token-based methods.
\textit{Third}, WM during LLM training methods 
embed WMs by training LLMs with specific triggers, knowledge distillation, reinforcement learning, or by modifying the LLM architecture \cite{xu2024learning,gu2023learnability,sun2022coprotector, sun2023codemark, baldassini2024cross}. These methods offer inherent robustness but are often associated with implementation complexity and increased computational overheads. \textit{In addition}, 
several miscellaneous WM approaches, such as mixed methods, multiple-output generation with selection, or prompt-based techniques, offer enhanced robustness and adaptability \cite{giboulot2024watermax,pang2024adaptive,huang2024waterpool,zhu2024duwak}. However, these methods come with trade-offs, including higher computational costs and potential vulnerabilities to watermark removal attacks.

\subsection{WM IP Checkers}

After watermarking the LLM outputs, the clouds can use IP checkers to determine 
whether  an adversary model has utilized watermarked outputs to replicate the behavior of the cloud-hosted model. This is typically done by checking whether the generated text from the model is watermarked.  IP checkers are generally divided into two main approaches. First, some IP checkers categorize each token as green or red and then apply statistical tests, such as z-scores, p-values, or Jensen-Shannon divergence, to assess whether the proportion of green tokens exceeds a specified threshold. If this threshold is met, the text is flagged as watermarked \cite{kirchenbauer2023watermark,liu2024semanticinvariantrobustwatermark,zhao2024provable}.  Second, other IP checkers assess the alignment between generated tokens and pseudo-random sequence, checking whether each token matches the corresponding value in the  sequence. Depending on a WM technique, the cloud will  apply its specific detection function to assess the output. If the statistical tests or alignment scores exceed certain thresholds, the output is considered watermarked. If the proportion of detected watermarks is sufficiently high, the model may be flagged as potentially stolen.

\subsection{Characteristics of an Effective WM}


A variety of WMs in LLMs have been introduced, while effective WMs in LLMs share key characteristics:
\textit{First,} WMs should minimally impact text quality, ensuring they are not noticeable to users. Quality can be assessed using metrics like perplexity, BLEU score, or accuracy on tasks like MMLU or sentiment classification \cite{hendrycks2021measuringmassivemultitasklanguage}.
\textit{Second,} WMs must be easily detectable by IP checkers, enabling reliable origin tracing with high detection rates.
\textit{Third,} WMs should be robust to WM removal attacks that attempt to remove watermarked patterns to bypass detection. By satisfying these characteristics, WMs  in LLMs can effectively balance security, model utility, and usability in real-world deployments.

\subsection{Attacks against WMs  }


\textbf{WM Removal Attacks \cite{zhang2023watermarks, krishna2024paraphrasing, pan2024markllm, pang2024attacking}.}  The attacker's goal is to remove embedded signatures and patterns from a watermarked text using paraphrasing techniques such as synonym replacement, sentence reordering, or sentence matching to minimize detectable traces of the WM. These attacks aim to generate semantically equivalent but syntactically modified text to evade WM detection mechanisms.  In addition, some attacks \cite{zhang2023watermarks} propose leveraging a high-quality oracle model to preserve output quality while removing the WM, making detection even more challenging.  However, these approaches often require additional training to effectively rephrase sentences or entire paragraphs, which can introduce subtle shifts in semantic meaning, degrade text quality, and significantly increase computational complexity.



\textbf{Spoofing Attacks \cite{pang2024attacking, jovanovic2024watermark, zhang2024large, sadasivan2023can,gu2023learnability}.} The attacker aims to produce text that is falsely detected as watermarked, thereby nullifying the value of WMs \cite{pang2024attacking}.  These attacks undermine the integrity of the WM by making it difficult to distinguish between legitimate watermarked content and counterfeited outputs. These approaches may harm the reputation of model providers, for example, if inappropriate or harmful text contents are falsely attributed to them. However, these attacks work only with certain types of WMs, and to be successful, they typically require a large number of queries to collect sufficient watermarked text for estimating the WM's distribution and reconstructing the WM rules, associating with computational overheads.



\begin{figure}[t]
      \centering
    \includegraphics[scale=0.1056]{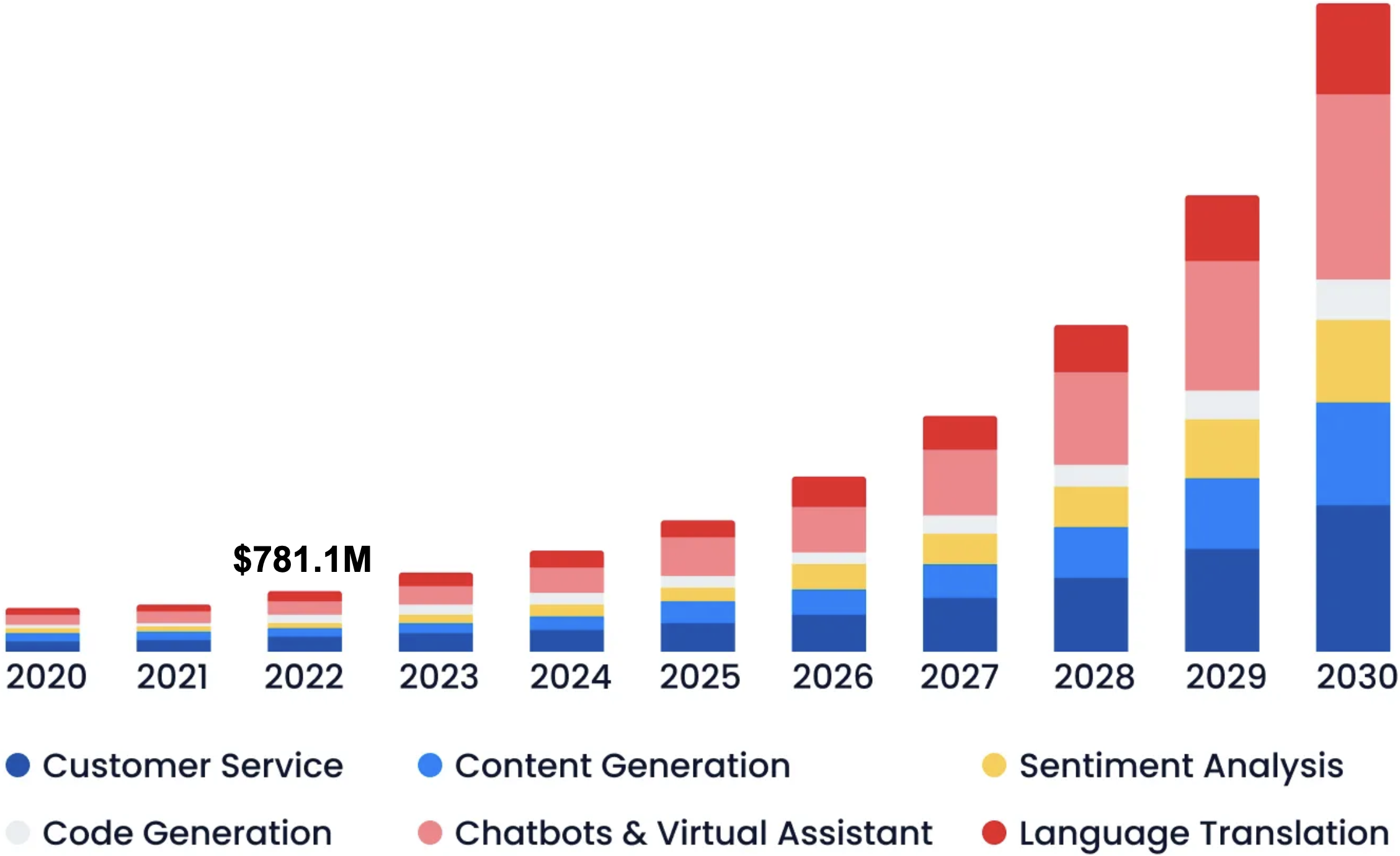} 
      \caption{US LLM Market Trend from 2020 to 2030 \cite{industry}. }  
      \label{fig:markettrend}
\end{figure}

\begin{table*}[t!]     
\centering
\caption{Real-world Services and Applications of LLMs. }
\begin{tabular}{|p{0.10\linewidth}|p{0.065\linewidth}|p{0.77\linewidth}|}
\hline
\textbf{API Name} &  \textbf{Provider} & \textbf{Real-world Services and Applications} \\ \hline
\hline
ChatGPT and GPT family  \cite{OpenAI} & OpenAI & 
\textbf{Salesforce} integrated its AI with OpenAI for CRM tools \cite{salesforce2023einstein}. 

\textbf{Duolingo}   introduced Duolingo Max powered by GPT-4 for better learning experience \cite{duolingo2023max}. 

\textbf{Instacart} added ChatGPT to its Ask Instacart chatbot for better customer interaction \cite{wsj2023instacart}. 

\textbf{Expedia} powered its chatbots by ChatGPT for comprehensive services \cite{expedia2023chatgpt}. 

\textbf{Morgan Stanley} introduced AskResearchGPT for staff support \cite{morganstanley2023openai}. 

\textbf{Stripe} uses GPT-4 to improve user experience and fight fraud \cite{stripe2023openai}.\\
\hline
Microsoft Azure Language \cite{MicrosoftAzureAI} & \small{Microsoft} Azure & 
\textbf{AT\&T} leveraged Azure OpenAI to automate processes and improve customer experiences \cite{MicrosoftAzureAI}.

\textbf{Vodafone} improved services with TOBi AI for personalized, multi-channel support \cite{azure2023openai}.

\textbf{Volvo Group} used AI for document translation to streamline invoice \& claims processing\cite{azure2023openai}.

\textbf{TotalEnergies} launched Microsoft Copilot for an AI chat solution using internal data \cite{microsoft2024ai}. 

\textbf{Providence} created ProvARIA based on Azure OpenAI to direct patient care messages \cite{microsoft2024ai}.

\textbf{Grupo Bimbo} developed a copilot for employee queries on  policies and risk management \cite{microsoft2024ai}.
\\ 
\hline
LLaMA family \cite{HuggingFaceMetaLLaMA} & Meta & 
\textbf{Zoom} utilized LLaMA-2 to create its AI Companion for meeting summaries, next-step highlights, and presentation tips \cite{meta2024llama}.

\textbf{Mathpresso} built MathGPT with LLaMA for personalized math learning \cite{meta2024llama}. 

\textbf{Goldman Sachs} used LLaMA to improve service, document review,  code generation\cite{meta2024llama}. 

\textbf{Nomura Holdings} used LLaMA to enhance customer service and document analysis\cite{meta2024llama}.

\textbf{EPFL} developed Meditron with LLaMA-2 to aid clinical decision-making and diagnosis\cite{meta2024llama}. 
\\ 
\hline
Google Cloud AI-Language (Gemini, Bard, PaLM, etc.) \cite{GoogleCloudNLP} & Google & 
\textbf{Lilt} integrated Google Translate into its platform for advanced translation capabilities \cite{lilt2024googletranslate}. 

\textbf{Wipro} with Google Cloud  created the Gemini Experience Zone for AI experimentation \cite{wipro2024news}. 

\textbf{General Motors} deployed a virtual assistant using Google Cloud's conversational AI for better intent recognition \cite{googlecloud2024aiusecases}.

\textbf{Gojek} launched "Dira by GoTo AI", an  voice assistant for GoPay service\cite{googlecloud2024aiusecases}.

\textbf{GroupBy} built a Search and Discovery Platform using Vertex AI to boost retail sales\cite{googlecloud2024aiusecases}.

\textbf{Motorola  AI} leveraged Gemini for conversation summaries and search\cite{googlecloud2024aiusecases}.
\\
\hline
Amazon Comprehend \cite{satyanarayana2020sentimental}  & Amazon AWS & 
\textbf{Vibes} used the platform for phrase extraction, sentiment analysis, and topic modeling \cite{aws2024comprehend}.

\textbf{Gallup} applied Targeted Sentiment to improve survey response reporting\cite{aws2024comprehend}.

\textbf{Zillow} built speech analytics to enhance customer and service support\cite{aws2024comprehend}.

\textbf{Siemens} leveraged AWS LLMs  to automate its surveying program\cite{aws2024comprehend}.
\\
\hline
IBM Watson NLU \cite{IBMNLP} & IBM Watson &
\textbf{SherloQ} used IBM Watson NLU to analyze search behavior and optimize ads \cite{ibm2024casestudies}. 

\textbf{University of Auckland} deployed IBM Watson Assistant based on a conversational AI solution for handling repetitive questions \cite{ibm2024casestudies}.  

\textbf{Mushi Lab} created Clearscope to optimize content with  insights \& recommendations \cite{ibm2024casestudies}.

\textbf{Global-Regulation} built a global law search engine for easier legal information access\cite{ibm2024casestudies}.
\\
\hline
Claude \cite{AnthropicClaude} & \small{Anthropic} & 
\textbf{Notion} integrated Claude for content generation, summarization, and workflow automation \cite{anthropic2024customers}.

\textbf{Decagon} utilized  Claude for personalized, 24/7 customer service \cite{anthropic2024customers}.

\textbf{GitLab} employed Claude to enhance coding assistance \&  software development\cite{anthropic2024customers}.

\textbf{Headstart} leveraged Claude to accelerate their software development process\cite{anthropic2024customers}.
\\
\hline
\end{tabular}
\label{table:llm_application}
\end{table*}

\section{Exploring Effectiveness of WMs in LLMs}

In this section, we focus on addressing the key question of this study: \textbf{\textit{Are WMs in LLMs Ready for Deployment?}} To provide a comprehensive answer, we 
first investigate the integration of  WMs in LLM  productions and discuss the existing and potential challenges of adapting them for practical use. In addition, we conduct an in-depth evaluation of the effectiveness of WMs across different LLMs by introducing a novel cross-model IP classifier and assessing how WMs impact the model utility of LLMs and their downstream tasks. By analyzing these aspects, we seek to determine whether WMs can be implemented without noticeably compromising the model utility of LLMs, which is a critical consideration in determining whether WMs are genuinely ready for widespread deployment.



\subsection{WMs in LLM Productions and Challenges}

LLMs have gained significant attention recently, with leading companies developing and deploying them in their products and services. 
Table \ref{table:llm_application}   illustrates real-world applications of various LLMs across industries. It highlights how these companies have integrated AI technologies like OpenAI's GPT, Microsoft's Azure, Meta's LLaMA, Google Cloud AI, Amazon Comprehend, and Anthropic's Claude to enhance their services. These examples demonstrate the broad impact of LLMs in areas such as customer service, content optimization, document processing, and software development, driving innovation and efficiency across sectors. In addition, 
as shown in Fig.~\ref{fig:markettrend}, the market for LLM production grew from 2020 to 2022, reaching $781.1$ million, with substantial growth expected by 2030, highlighting the increasing importance of LLMs across various industries.





Given the increasing trend in the LLM market, along with potential risks associated with their deployment in real-world scenarios, as discussed in Section \ref{sec: LLM and risks}, it is essential to utilize WMs to protect IP of LLM outputs.  
However,  the adoption of WMs in LLM productions remains limited.  
For instance, OpenAI has developed a WM system for ChatGPT but has opted not to implement it due to concerns over ambiguous potential penalties for users and feedback suggesting that service usage might decline if WMs were applied \cite{chatgptWM}. 
The uncertainty from providers and users around adopting WMs in LLMs  stems from a lack of understanding of WM effectiveness and how different LLMs interact with IP checkers for a given WM.

However, uncovering the true effectiveness of WMs in LLMs and such interactions presents several challenges.  
\textit{First,} while LLMs and WMs are widely used, the impact of WMs on  output quality and IP protection remains unclear, particularly regarding how WMs affect model utility and detection by IP checkers.
 \textit{Second,} when a cloud service receives outputs from different LLMs, both with and without WMs, how the cloud classifies which model generated the output and how WMs support this classification.
\textit{Third,} it is unclear which WM should be applied to a given LLM and why. Companies need clear and evidence-based guidelines for selecting the most effective WM strategies to ensure optimal performance and user experience.
\textit{Fourth,} there is a lack of standardized metrics to evaluate how WMs impact LLM utility, making it difficult to assess their effectiveness across different LLMs and tasks.  
\textit{Finally,} it is essential to assess model utility and the quality of LLM outputs on downstream tasks after watermarking to ensure that WMs do not degrade performance. 
Addressing these challenges is vital for successfully deploying WMs in real-world LLM applications, ensuring a balance between IP protection and model utility.




\subsection{Native IP Classifier}
To understand the effectiveness of WMs on LLMs in terms of model utility and IP checker performance,  a straightforward approach is to apply the specific IP checker tailored to the  WM on the LLM and compare the utility and quality of the text before and after being watermarked. 
The key idea behind the IP checker is to evaluate the statistical differences between original and watermarked outputs, flagging the output as watermarked if the differences surpass a set threshold.  
 In other words, the IP checker acts as a classifier to distinguish between the two outputs.

 \begin{figure}[t]
      \centering    
      \includegraphics[scale=0.043]{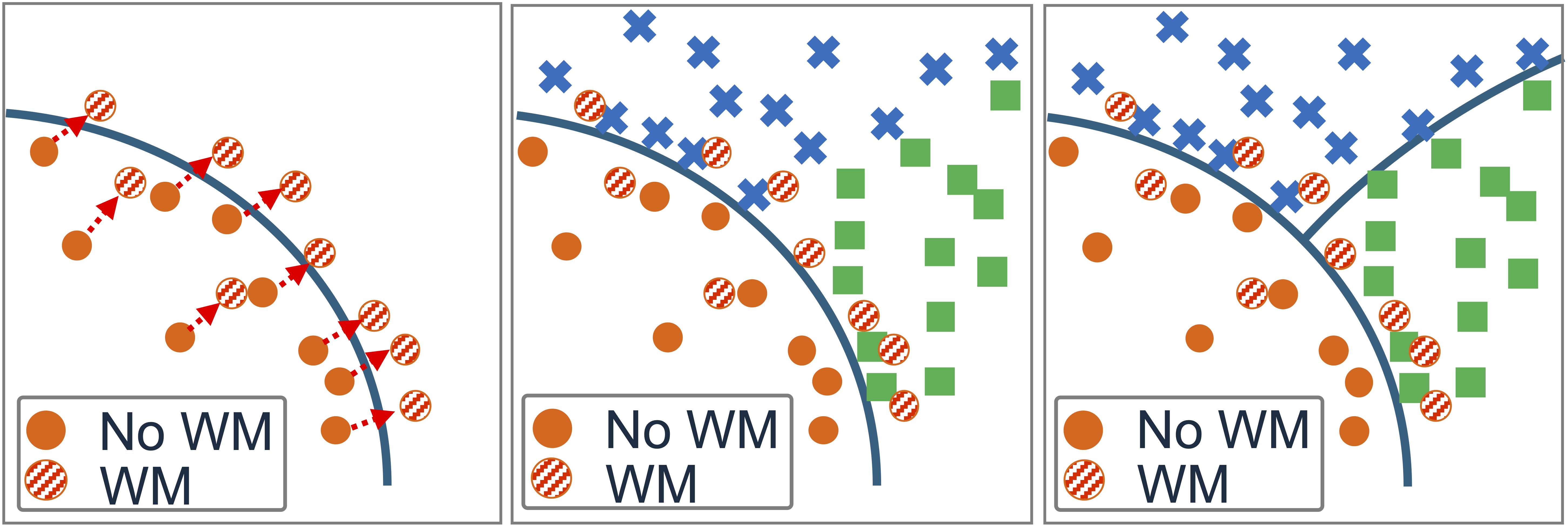} 
      \caption{Native IP checker works well in simple scenarios (left), but struggles to differentiate between outputs from multiple LLMs (center), highlighting the need for a cross-model IP checker (right).} 
      \label{fig:decision_boundary}
 \end{figure}

Despite its high performance, applying a native IP checker independently to each WM and LLM may be insufficient as it may not effectively capture the true impact and effectiveness of the WM.  In practice, a cloud provider may not know the exact source of the output, as multiple LLMs could be involved. Watermarked outputs from one LLM may closely resemble the original outputs from another, due to distribution shifts caused by the WM (Fig.~\ref{fig:decision_boundary}). Therefore, applying the IP checker in isolation to each LLM may yield misleading results. To more accurately assess WM effectiveness,  it is complemented to adopt a broader approach that evaluates the WM across multiple LLMs. 
  By introducing a cross-model IP classifier,  we can more reliably differentiate between the original and watermarked texts across different LLMs. This approach ensures  a cross-model IP classifier should be used. This approach ensures reliable differentiation between original and watermarked texts across multiple LLMs, better reflecting real-world conditions.





\subsection{Proposed Cross-Model IP Classifier}

  In this section, we present the following components: 1) a novel cross-model IP classifier to systematically discover the interplay between WMs and different LLMs,  2) a selection of evaluation metrics for IP assessment, 3) the impact of WMs on LLMs' utility, and 4) based on comprehensive experiments, we provide an insightful understanding of advantages and constraints of different WMs in LLMs. 

The key concept behind our cross-model IP classifier is to distinguish between the output distributions of different LLMs, both with and without WMs, reducing ambiguity in the IP classifier's decision boundary. This classifier systematically connect WMs and different LLMs through a multi-class model functioning as an IP checker.





\begin{figure}[t]
\centering
\subfloat[No WM across LLMs]{\label{fig:subnets-aa}\includegraphics[scale=0.033]{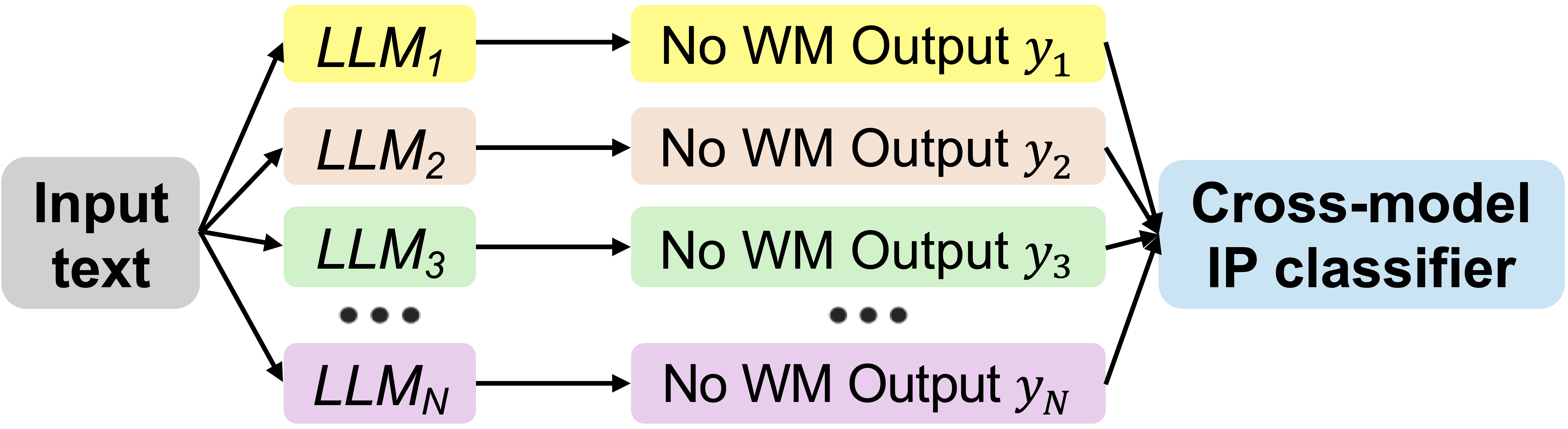}}\hfill
\subfloat[A WM for only one LLM at a time]{\label{fig:subnets-bb}\includegraphics[scale=0.033]{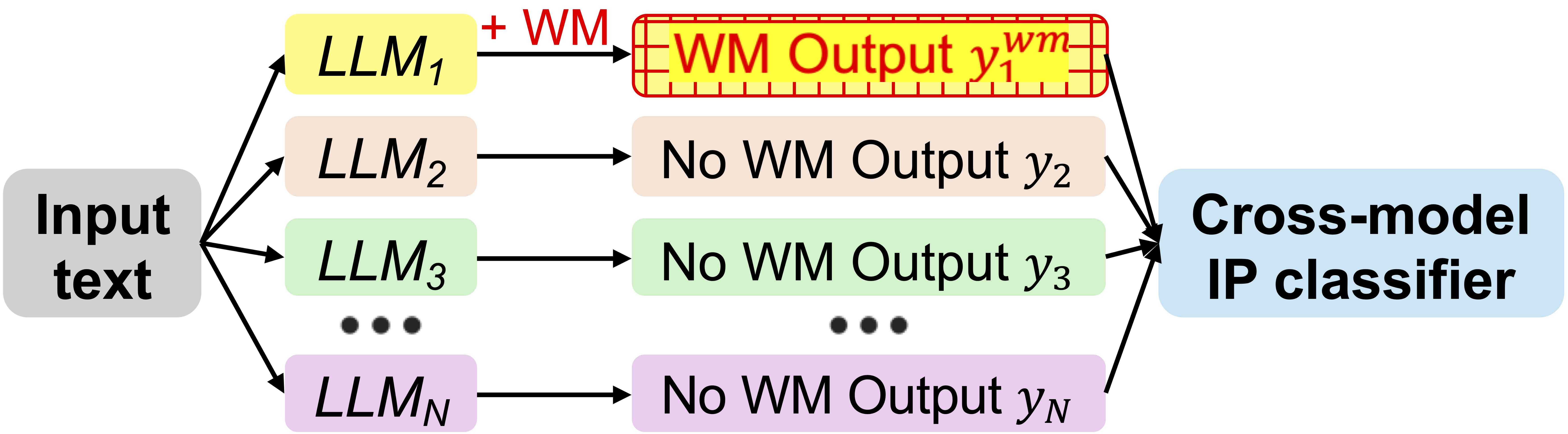}}\hfill 
\subfloat[A WM for all LLMs]{\label{fig:subnets-cc}\includegraphics[scale=0.033]{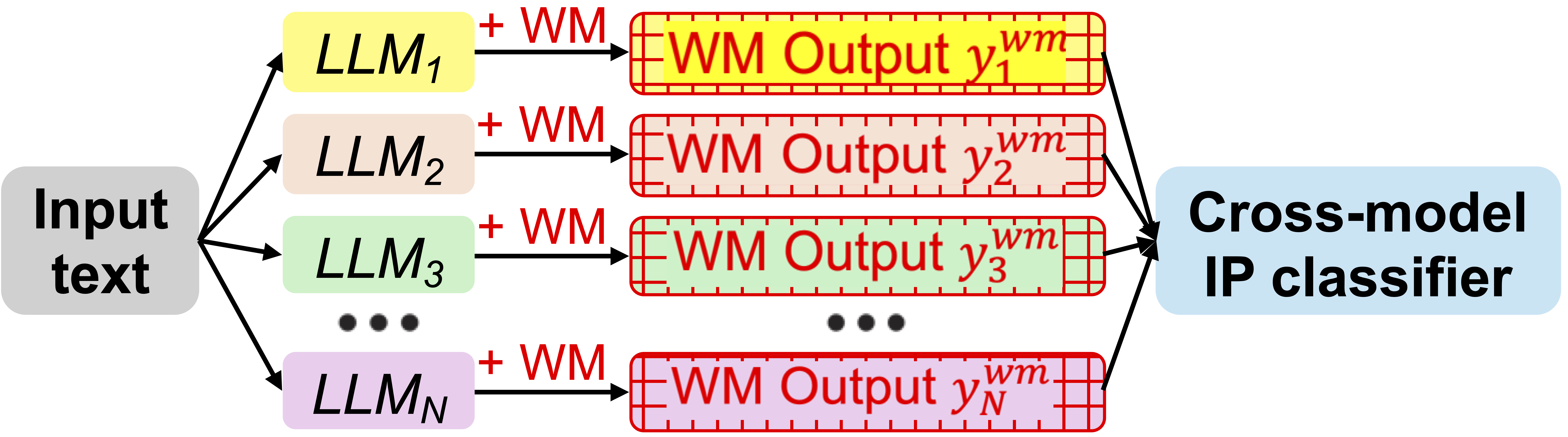}}
\vspace{-5pt}
\caption{Our Proposed Cross-model IP Classifier.}
\label{fig:setting}  
\end{figure}

Given a set of $N$ LLM models  $\{LLM_1, LLM_2, \cdots, LLM_N \}$ and an input text $x \in D$, we generate two sets of outputs, including outputs $y$ without WMs and outputs $y^{wm}$ using WMs.  
The IP classifier is designed to assess whether $x$ belongs to which LLM. In our cross-model classifier, we consider three scenarios.  \textit{First,} we use the outputs $y_i$ from all the $LLM_i$ ($i \in [1,N]$) as inputs for the classifier $f$ (Fig.~\ref{fig:setting}a). This setting discloses how different LLMs distinguish their outputs compared with others, highlighting the uniqueness of each LLM system. \textit{Second,} we apply WMs into one LLM at a time to thoroughly investigate the relationship between each LLM and WM without being affected by the results of other LLMs   (Fig.~\ref{fig:setting}b). \textit{Third,} we unify the evaluations to uncover the overall correlation among WMs and LLMs by applying each WM to all LLMs (Fig.~\ref{fig:setting}c). 

Next, we will conduct experiments to evaluate our proposed method in assessing the effectiveness of WMs on LLMs, analyze the results, and discuss the limitations of existing approaches, along with potential solutions.





\begin{figure*}[t] 
\centering
\subfloat[KGW]{\label{fig:subnets-a}\includegraphics[scale=0.305]{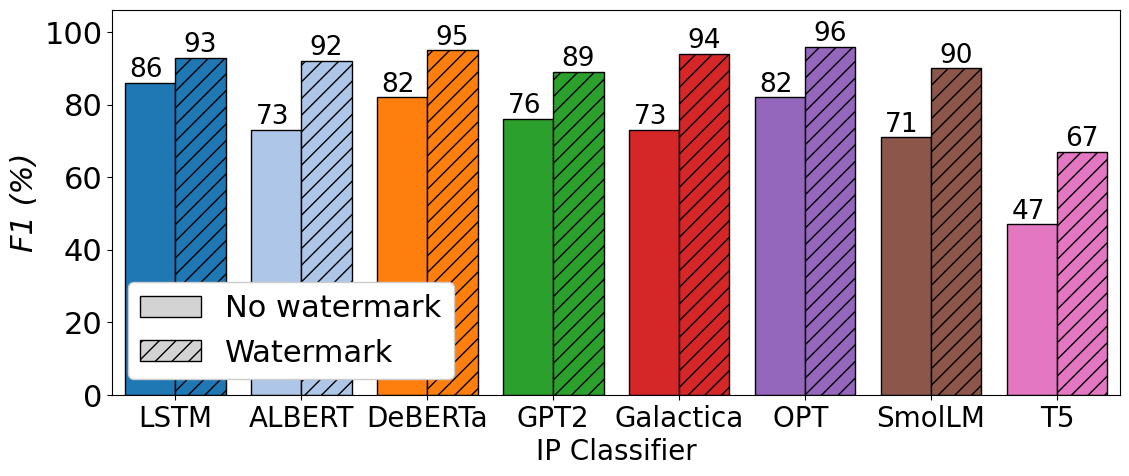}}\hfill
\subfloat[EXP]{\label{fig:subnets-b}\includegraphics[scale=0.305]{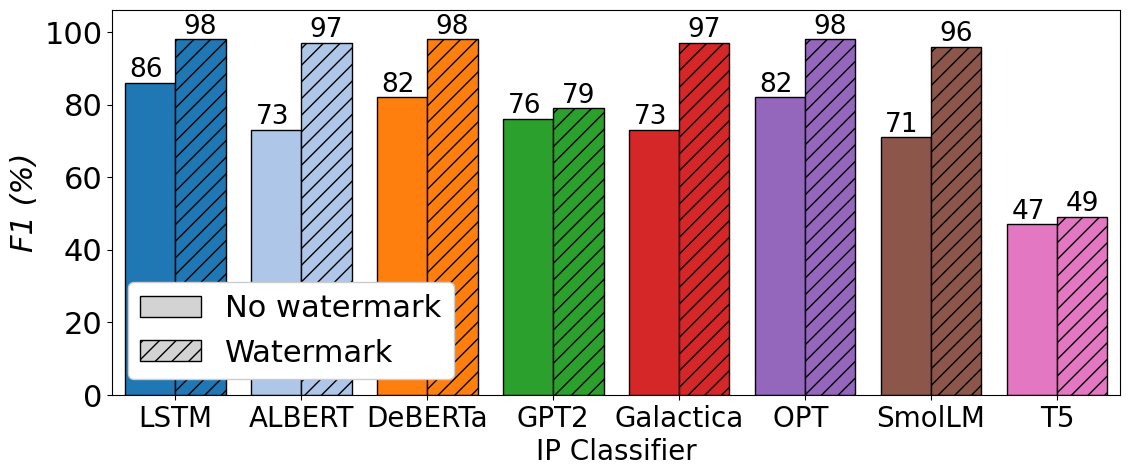}}\hfill 
\subfloat[SIR]{\label{fig:subnets-c}\includegraphics[scale=0.305]{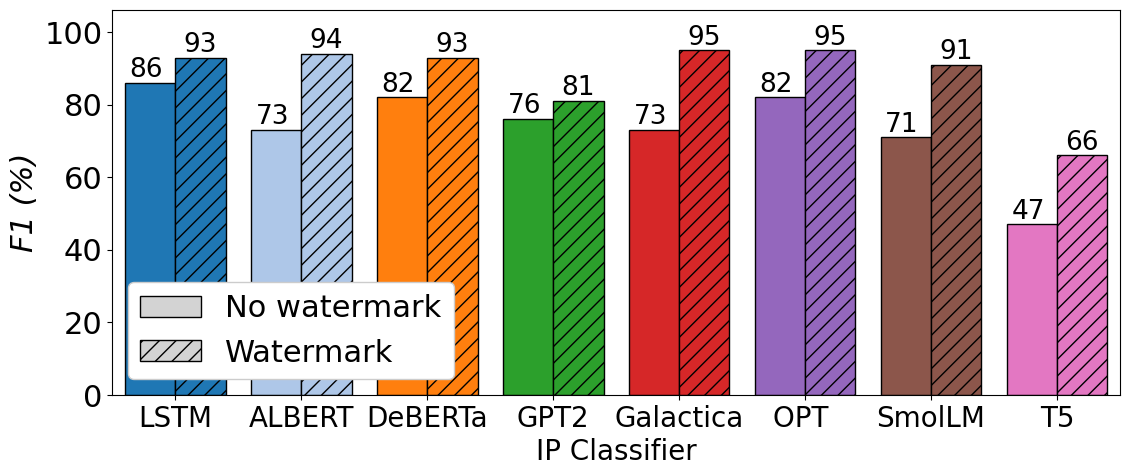}}\hfill 
\subfloat[SemStamp]{\label{fig:subnets-d}\includegraphics[scale=0.305]{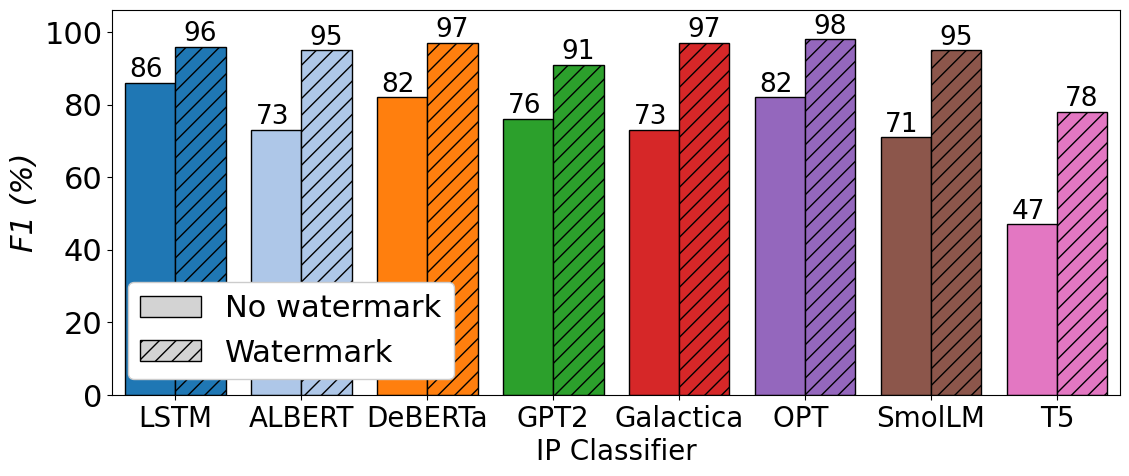}}\hfill 
\caption{$F1$ scores as a function of different IP classifiers for different WMs, compared with No WM settings.
} 
\label{fig:IPChecker}  
\end{figure*}

\subsection{Experiment Settings}

\subsubsection{Baselines} 
We carry out the validation on a wide range of WMs,  LLMs, and IP classifiers. \textit{For LLMs,} we use four state-of-the-art (SOTA) pre-trained LLMs, including LLaMA-2 with 7B parameters \cite{touvron2023llama2openfoundation}, Mistral with 7B parameters\cite{jiang2023mistral7b}, OPT with 6.7B parameters\cite{zhang2022opt} and Falcon with 7B parameters\cite{almazrouei2023falcon}. \textit{For WMs,} we use seven SOTA WM schemes, including 1) KGW \cite{kirchenbauer2023watermark}, 2) EXP \cite{kuditipudi2023robust}, 3) SIR \cite{liu2024semanticinvariantrobustwatermark}, 4) SemStamp \cite{hou2023semstamp}, 5) Unigram \cite{zhao2024provable}, 6) Adaptive \cite{liu2024adaptive}, and 7) UPV \cite{liu2023unforgeable}. These WMs are representative WM schemes  outlined in  Section \ref{sec:WMtaxonomy}, with further details provided in  Table \ref{table:7WMs} (Appendix). 

\textit{For IP classifiers}, we explore a range of options, including \textit{1)} a non-transformer LSTM model \cite{hochreiter1997long}; \textit{2)} an encoder-based transformer model \textit{ALBERT} \cite{DBLP:journals/corr/abs-1909-11942}, with 11.8M parameters; \textit{3)} another larger encoder-based transformer \textit{DeBERTa} \cite{he2021deberta} with 86M parameters; \textit{4)} a decoder-based transformer \textit{GPT-2} \cite{radford2019language} with 125M parameters; \textit{5)} a decoder-only model \textit{Galactica} \cite{taylor2022galactica} with 125M parameters; \textit{6)} a small version of \textit{OPT} \cite{zhang2022opt} with 125M parameters; \textit{7)} the recently developed \textit{SmolLM} \cite{huggingface2024smollm} with 135M parameters; and \textit{8)} an encoder-decoder model \textit{T5} \cite{raffel2020exploring},  with 60.5M parameters. These models were selected based on the following criteria: 1)  inclusion of non-transformer and transformer-based architectures, 2) within transformer-based models, coverage of all  major categories outlined in Section \ref{sec: LLM and risks}, and 3) consideration of model complexity as determined by parameter size. 

To further assess the effectiveness of WMs, we test them against SOTA attacks, including 1) a WM removal attack \cite{zhang2023watermarks}, referred to as \textit{WMremoval}, which removes WMs by perturbing quality-preserving WM outputs to produce high-quality non-WMed outputs; 
2) \textit{Dipper} \cite{krishna2024paraphrasing}, which paraphrases paragraphs  through content reordering and lexical changes without appreciably modifying the text semantics, and 3) \textit{Substitution} attack \cite{pan2024markllm}, which randomly substitutes target words with their synonyms, considering the surrounding context and using WordNet \cite{miller1995wordnet}. We adopt the default settings reported in these papers.

\subsubsection{Datasets and Evaluation Tasks}

We randomly select $10,000$ training samples and $2,000$ test samples from the benchmark C4 dataset \cite{dodge2021documentinglargewebtextcorpora}. Each sample, a news-like string, is truncated to $200$  tokens as a prompt, with the next $200$  tokens used as the ``baseline'' completion. Tasks include text generation as the primary task and massive multi-task language understanding (MMLU) \cite{hendrycks2021measuringmassivemultitasklanguage} as a downstream task, assessing the impact of WMs on LLM utility. MMLU has $57$ subjects in STEM, humanities, and social sciences, covering various difficulty levels to evaluate world knowledge and problem-solving skills.

\subsubsection{Evaluation Metrics}
We treat the IP evaluation across different LLMs as a multi-class classification problem, using the $F1$ score for IP classifiers. To measure the impacts of WMs, we calculate the change in $F1$ score, denoted as $F1_{change}$, by computing a relative change in $F1$ between scenarios with and without WMs.  The metrics are calculated as follow:
\begin{align}
  &   F1 = 2 \times \frac{precision \times recall}{precision + recall} \\
  & F1_{change}= \frac{F1^{WM}-F1^{NW}}{F1^{NW}}
\end{align}
 where $precision$ is defined as the number of true positive results divided by the total number of all samples predicted by the model, including incorrect predictions, and $recall$ is the number of true positive results divided by the total number of all actual instances. $WM$ and $NW$ indicates the score from WM and no WM settings, respectively.




In addition, to evaluate the model utility comprehensively, we consider 1)   Perplexity for the text generation task,  measuring how well a model predicts next tokens \cite{mikolov2011empirical} and 2) Average accuracy across topics for the MMLU downstream task. 
To further assess text quality, we provide qualitative examples comparing a prompt, the original output without WMs, and corresponding outputs with different WMs to offer insights into WM effectiveness.

\subsection{Experimental Results}

Under our IP classifier, an LLM is considered effective if it achieves a high F1 score (derived from settings a and c in Fig.~\ref{fig:setting}, where all LLMs are under the same condition of either with or without watermark), indicating the uniqueness of the LLM's outputs and enabling easy differentiation from other LLMs. This capability also allows cloud providers to determine whether an output stems from their model, thereby effectively protecting their IP.  Meanwhile, a WM scheme is deemed effective if (1) $F1_{change}$, derived from setting b in Fig.~\ref{fig:setting},  is significantly high, facilitating better detection compared to a scenario where no WM is applied, and (2) it has minimal impact on the model's utility of the main task (i.e., maintaining low perplexity in text generation) and of downstream tasks (i.e., high average accuracy of MMLU). This ensures that WMs enhance the uniqueness of the LLM's outputs without adversely affecting the model's overall utility.


\begin{figure}[t]
\centering
\subfloat[T5]{\label{fig:subnets-t5}\includegraphics[scale=0.265]{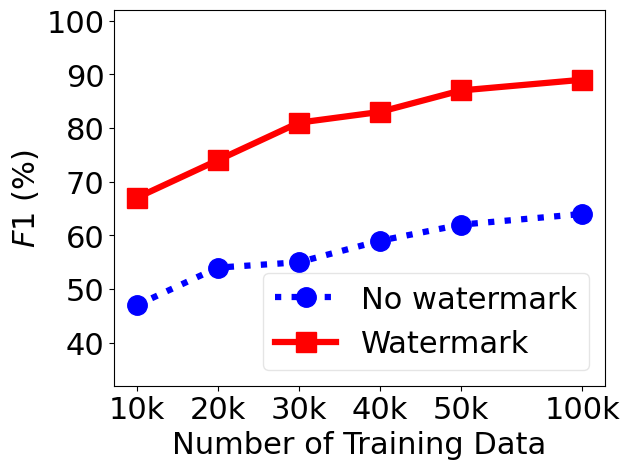}}\hfill
\subfloat[DeBERTa]{\label{fig:subnets-deberta}\includegraphics[scale=0.265]{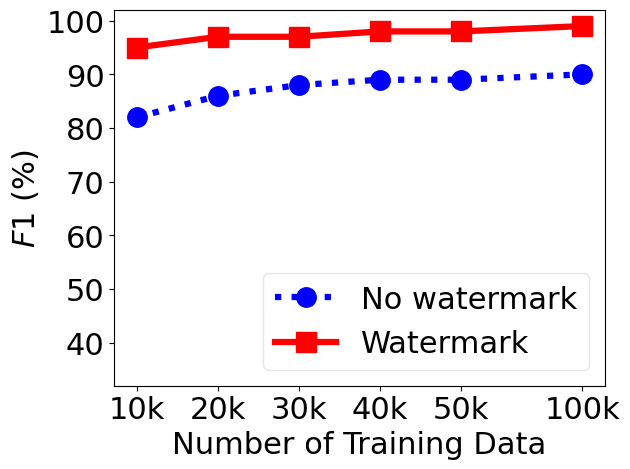}}
\caption{$F1$ score as a function of number of training data applied on KGW and LLaMA-2.}
\label{fig:T5trainning}  
\end{figure}

\subsubsection{Uniqueness of LLM Outputs}
Figs.~\ref{fig:IPChecker} and \ref{fig:IPChecker_addition} (Appendix)
illustrate $F1$ scores of different IP classifiers in distinguishing outputs from different LLMs. 
Without WMs, these IP classifiers achieve impressive $F1$ scores of over $70\%$ across various types. The LSTM outperforms others, achieving an F1 score of $86\%$. These results underscore the uniqueness of LLM outputs, allowing for reliable identification of the originating LLM based on the observed outputs. Notably, the T5 IP classifier performs poorly without watermarks, achieving only 47\%. Indeed, T5 is an encoder-decoder model that is more suitable for tasks such as translation and completion. In Fig.~\ref{fig:T5trainning}, the poor results  stem from insufficient training data. With increased numbers of training samples, the classifier's $F1$ score improves, reaching $64\%$ with $100,000$ training samples. Similar improvements are observed in other models like DeBERTa, with an 8\% increase as training samples grow.   

\subsubsection{Effectiveness of WMs on LLMs} 
\noindent \textbf{Impacts of WMs on Classifying LLM Outputs. } 
As shown in Figs.~\ref{fig:IPChecker} and \ref{fig:T5trainning}, WMs can significantly boost the performance of LLM output classification.  We observe this phenomenon across WMs, LLMs, and IP classifiers. There is a significant improvement of  $2\%-31\%$, compared with the performance without WMs. In particular, ALBERT, T5, Galactica, and SmolLM classifiers demonstrate substantial improvements when WMs are applied, with an increase of  $19\%-31\%$ in $F1$ scores. It is worth noting that all the classifiers achieve over $90\%$ $F1$ score with WMs. 
Notably, WM with EXP or Semstamp allows the OPT classifiers to reach an impressive $98\%$ $F1$ score.   
The enhanced performance in distinguishing LLM outputs is due to the unique patterns or identifiers created by WMs, which increase the distinctiveness of each model's output, making it easier to identify the source LLM.  Even with the T5 classifier, which performs poorly on non-watermarked text, shows significant improvement with WMs, increasing from $49\%$ to $78\%$.


The remarkable improvements in IP classifier performance highlight the effectiveness of WMs in differentiating outputs from various LLMs. This crucial correlation between LLMs and WMs allows clouds to  identify their outputs accurately, ensuring that proprietary content is  protected. By leveraging WMs, clouds can enhance  IP protection for their models and reduce unauthorized use. This capability is essential in competitive environments such as industry, where protecting unique outputs is critical for business success.



\begin{figure}[t]
\centering
\subfloat[ALBERT]{\label{fig:subnets-albert}\includegraphics[scale=0.215]{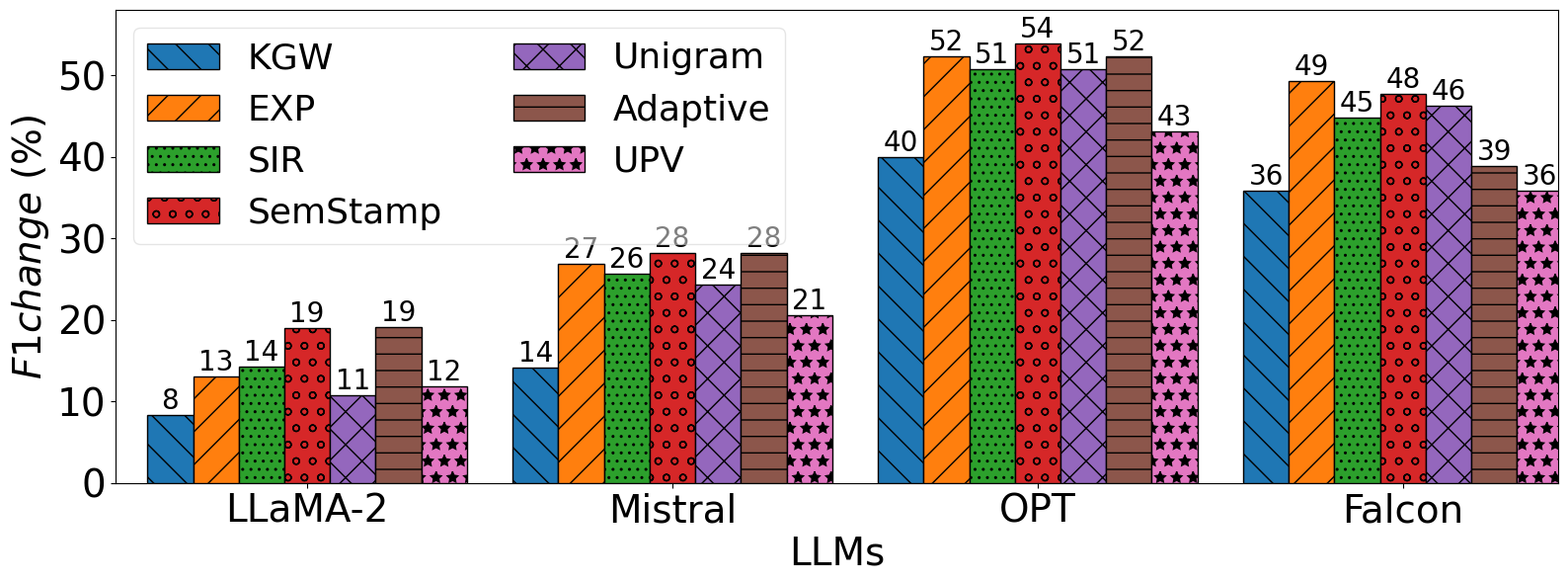}}\hfill
\subfloat[DeBERTa]{\label{fig:subnets-deberta2}\includegraphics[scale=0.215]{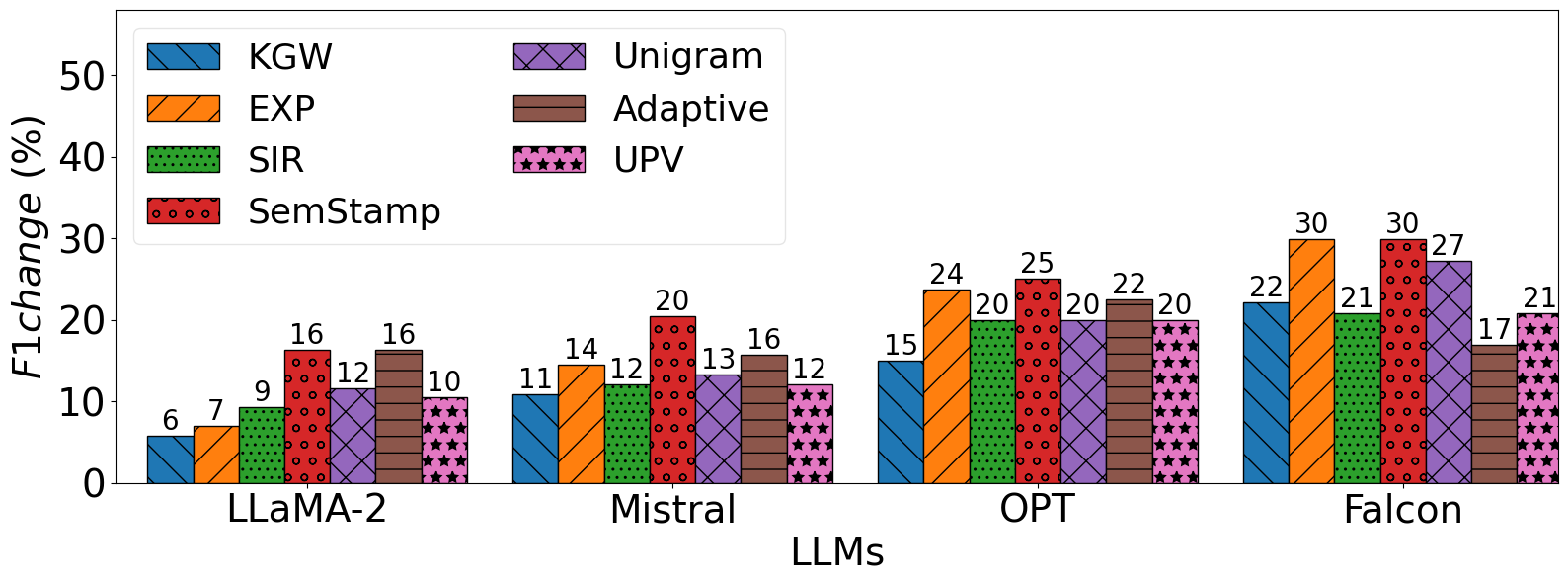}}
\caption{$F1_{change}$ across WMs, LLMs, and IP classifiers.}
\label{fig:LLM_WM}  
\end{figure}

\noindent  \textbf{Impacts of LLMs on IP Classifiers.}
Fig.~\ref{fig:LLM_WM} demonstrates the impact of different LLMs on IP classifiers. These results are obtained by applying WMs to one LLM at a time, with the relative change $F1_{change}$ indicating the impacts. When WMs are applied, all LLMs show positive values of $F1_{change}$, signifying an increase in their $F1$ scores. In this analysis, we present representative trends observed  in the ALBERT and DeBERTa classifiers. Both classifiers demonstrate a consistent pattern, showing   more significant changes for the OPT and Falcon models, compared with the LLaMA-2 and Mistral. This variation is  
attributed to LLaMA-2 and Mistral's adaptability and robustness, which make them more resistant to the effects of WMs. In contrast, OPT and Falcon are specifically designed for efficiency and scalability. 

\noindent  \textbf{Impacts of WMs on Model Utility of Text Generation Tasks.}
In addition to improving IP classifiers' ability to distinguish LLM outputs, it is crucial to assess how different WMs impact model utility, as this is a key characteristic of effective WMs.  
As shown in Table \ref{table:PPL_scores}, 
WMs lead to a noticeable increase in the perplexity of the generated text, indicating a negative effect on the model's utility in text generation.
KGW has the lowest impact on perplexity values across all LLMs, while SemStamp exhibits substantial changes in perplexity. 
For instance, with the OPT model, KGW slightly changes perplexity from $3.55$ to $3.87$. In contrast, SemStamp significantly increases perplexity to $76.91$ in this setting. 
The key reason for this difference lies in how each WM operates. To elaborate, KGW applies the WM during the logits generation process, modifying the text without significantly altering the model, which minimizes its impact on the LLMs. Meanwhile, SemStamp influences the sampling process by generating new sentences from a partitioned valid semantic space, resulting in a more noticeable effect on the generation process, compared with other baselines. It is worth noting that all types of WMs produce considerable impact on this text quality, urging for further investigation.


 \begin{figure}[t]
      \centering    
      \includegraphics[scale=0.3]{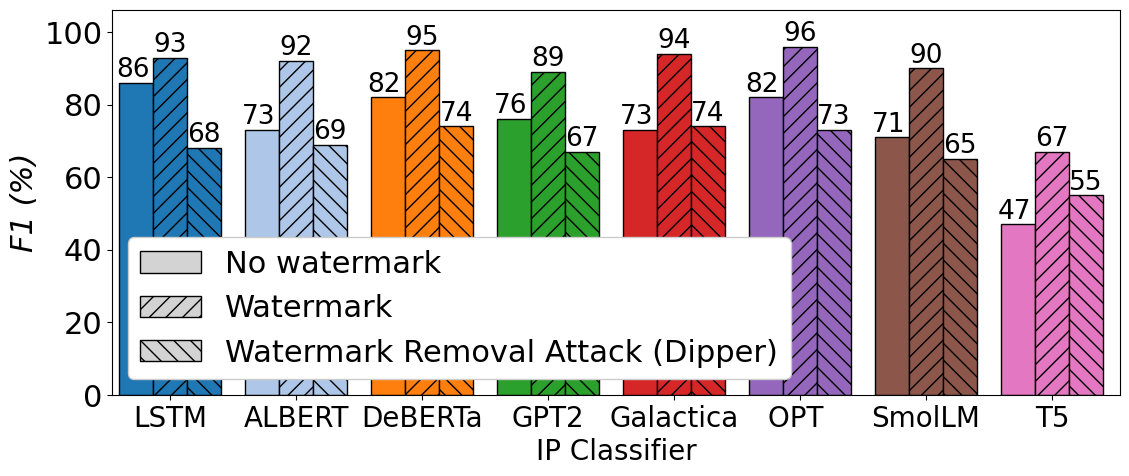} 
      \caption{$F1$ scores as a function of different IP classifiers for KGW WMs, compared with No WM and after WM Removal Attack of Dipper.} 
      \label{fig:Attack_KGW_Dipper}
 \end{figure}

\noindent  \textbf{Impacts of WMs on Model Utility of Downstream Tasks.}

We further conduct an in-depth analysis  in the MMLU question-answering task (Table \ref{table:MMLU}). Our observations reveal that majority of WMs have subtle impact on the average accuracy of MMLU. For instance, there are up to $2.3\%$ decreases in the accuracy across LLMs and WMs such as KGW, EXP, Unigram, and UPV compared with the outputs without WMs, indicating that watermarked LLM outputs still retain usefulness for classification tasks. However, WMs like SIR and Adaptive show more significant accuracy drops across all four LLMs. The primary reason for these poor results lies in our reliance on their publicly available pre-trained models (such as the semantic mapping model for Adaptive and the watermarked model for SIR) without fine-tuning them for our specific tasks. Limited time and computational resources prevented us from aligning these models with our experimental settings, which likely affected their performance. Additionally, SemStamp, which involves sentence-level sampling, is not ideal for MMLU tasks, where token-level generation is preferred.

\begin{figure*}[t]
\centering
\subfloat[LLaMA-2 + KGW]{\label{fig:subnets-1}\includegraphics[scale=0.265]{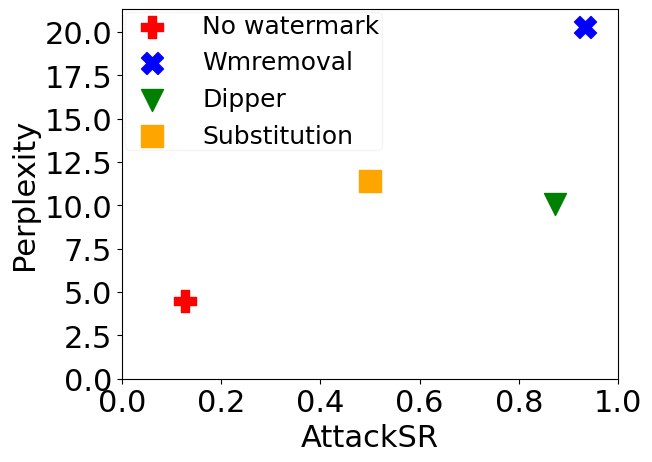}}\hfill
\subfloat[Mistral + EXP]{\label{fig:subnets-2}\includegraphics[scale=0.265]{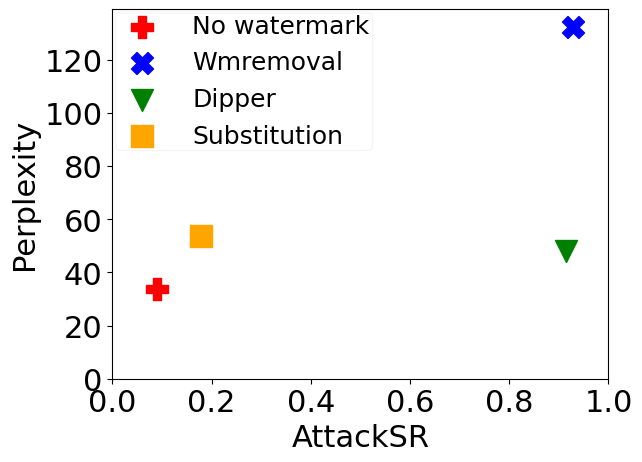}}\hfill
\subfloat[LLaMA-2 + SIR]{\label{fig:subnets-3}\includegraphics[scale=0.265]{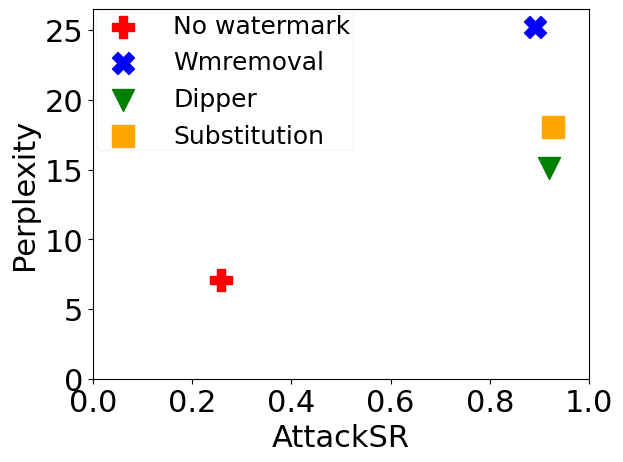}}\hfill
\subfloat[LLaMA-2 + SEM]{\label{fig:subnets-4}\includegraphics[scale=0.265]{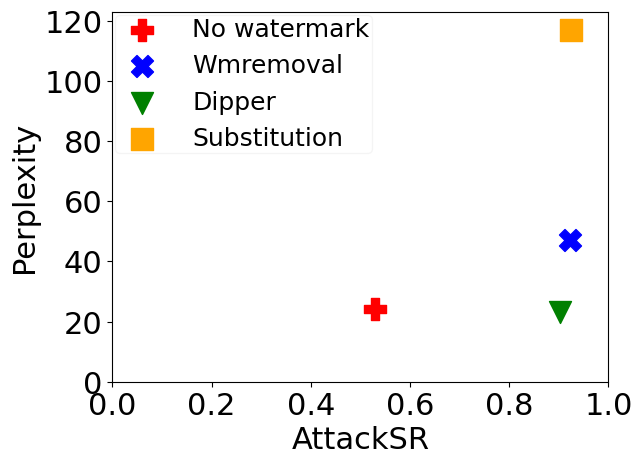}}\\
\subfloat[Mistral + KGW]{\label{fig:subnets-5}\includegraphics[scale=0.265]{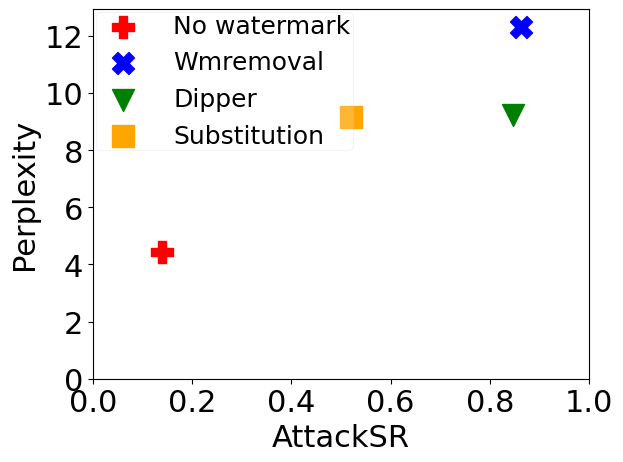}}\hfill
\subfloat[LLaMA-2 + EXP]{\label{fig:subnets-6}\includegraphics[scale=0.265]{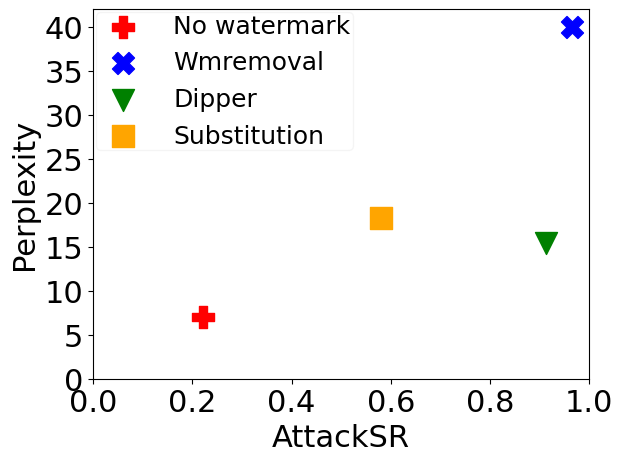}}\hfill
\subfloat[Mistral + SIR]{\label{fig:subnets-7}\includegraphics[scale=0.265]{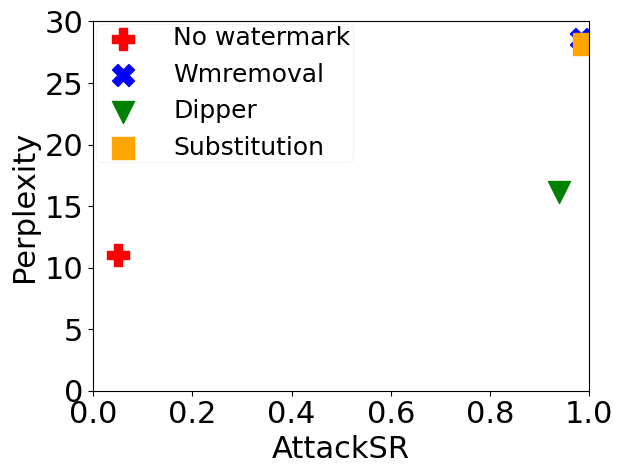}}\hfill
\subfloat[Mistral + SEM]{\label{fig:subnets-8}\includegraphics[scale=0.265]{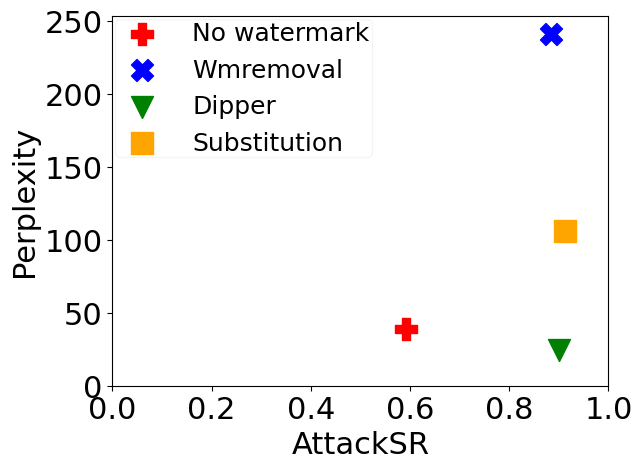}}
\caption{Attack success rate and Perplexity across different attacks and WMs with LLaMA-2 and Mistral models.}
\label{fig:compare_with_others}  
\end{figure*}

\noindent \textbf{WMs against Attacks.} We further examine how WMs resist against different  attacks, indicating the effectiveness and resilience of WMs on LLMs. We consider WM removal attack, Dipper, and substitution attacks. Fig.~\ref{fig:compare_with_others} shows that these attacks achieve high attack success rates, which indicates a low detection rate of the WM's detection function. However, it comes with the cost of a significant impact on model utility, as evidenced by a substantial increase in perplexity. For instance, WM removal attacks can raise the perplexity of a non-watermark baseline by 2 to 6 times. In the case of the Mistral model with Semstamp, perplexity increases dramatically from $39.31$ up to $241$. In addition, Fig.~\ref{fig:Attack_KGW_Dipper} demonstrates the impact of the Dipper attack on WM effectiveness in classifying LLM outputs. The results show a significant performance drop, highlighting the detrimental effects of the attack on both the effectiveness of WMs and the quality of LLM outputs.

\begin{table}[t]
\centering
\caption{Perplexity values of different WMs across LLMs.}
\resizebox{.50\textwidth}{!}{
\begin{tabular}{|p{1.4cm}|p{1.38cm}|p{1.15cm}|p{1.25cm}|p{1.2cm}|p{1.25cm}|}
\hline
\textbf{Setting} & \small{\textbf{LLaMA-2}} & \textbf{Mistral} & \textbf{OPT} & \textbf{Falcon} & \textbf{Average} \\ \hline
\textbf{NW} & 3.61 & 3.54 & 3.55 & 2.68 & 3.35 \\ \hline
\textbf{KGW} & 4.46 \textit{(24\%)} & 4.43 \textit{(25.1\%)} & 3.87 \textit{(9.0\%)} & 3.37 \textit{(25.7\%)} & 4.03 \textit{(20.6\%)} \\ \hline
\textbf{EXP} & 7.06 \textit{(96\%)} & 33.61 \textit{(849.4\%)} & 21.10 \textit{(494.4\%)} & 18.72 \textit{(598.5\%)} & 20.12 \textit{(501.6\%)} \\ \hline
\textbf{SIR} & 7.05 \textit{(95\%)} & 11.00 \textit{(210.7\%)} & 19.72 \textit{(455.5\%)} & 13.89 \textit{(418.3\%)} & 12.92 \textit{(286.1\%)} \\ \hline
\textbf{SemStamp} & 24.06 \textit{(566\%)} & 38.89 \textit{(998.6\%)} & 76.91 \textit{(2066.5\%)} & 7.52 \textit{(180.6\%)} & 36.85 \textit{(1001.5\%)} \\ \hline
\textbf{Unigram} & 5.50 \textit{(52\%)} & 8.68 \textit{(145.2\%)} & 13.71 \textit{(286.1\%)} & 10.71 \textit{(299.7\%)} & 9.65 \textit{(188.4\%)} \\ \hline
\textbf{Adaptive} & 15.77 \textit{(337\%)} & 12.97 \textit{(266.4\%)} & 11.24 \textit{(216.5\%)} & 5.47 \textit{(104.0\%)} & 11.36 \textit{(239.6\%)} \\ \hline
\textbf{UPV} & 5.81 \textit{(61\%)} & 4.39 \textit{(24.1\%)} & 4.69 \textit{(32.2\%)} & 3.45 \textit{(28.9\%)} & 4.59 \textit{(37.1\%)} \\ \hline
\end{tabular}}
\label{table:PPL_scores}
\end{table}

\begin{table}[t]
\centering
\caption{MMLU Accuracy (\%) of different LLMs with and without WMs.}
\begin{tabular}{|l|c|c|c|c|}
\hline
\textbf{Settings} & \textbf{LLaMA-2} & \textbf{Mistral} & \textbf{OPT} & \textbf{Falcon} \\ \hline
\textbf{No WM} & 46.7 & 58.9  & 24.8 & 27.2  \\ \hline
\textbf{KGW}  & 44.5  & 58.6  & 24.5  & 25.3  \\ \hline
\textbf{EXP}  & 45.9   & 58.7  & 24.8  & 24.9  \\ \hline
\textbf{SIR } & 27.9   & 20.9  & 18.3  & 19.8  \\ \hline
\textbf{Unigram}  & 46.6   & 58.1  & 23.8  & 25.3  \\ \hline
\textbf{Adaptive}  & 31.4   & 37.9  & 24.7  & 26.1  \\ \hline
\textbf{UPV}  & 46.7   & 57.9  & 24.7 & 25.6  \\ \hline
\end{tabular}
\label{table:MMLU}
\end{table}

\subsubsection{Semantic Preservation of Watermarked Outputs}
Furthermore, to qualitatively assess the quality of generated text under different WMs, we illustrate side-by-side examples of real prompts, original outputs from an LLaMA-2 model without WMs and with different WMs. 

In Table \ref{table:viz_LLM2_text}, we observe that watermarked text preserves the semantic meaning compared with the original outputs without WMs. Notably, KGW maintains high semantic similarities, as  the green-highlighted phrases show. For instance, key phrases such as ``are suing the college," ``medical laboratory technician program," and ``filed a class-action lawsuit" are preserved almost identically in both the KGW output and the original, demonstrating KGW's ability to maintain the integrity of the original semantic content. This is further supported by the low perplexity of $2.75$, close to the non-WM perplexity of $2.32$, indicating minimal degradation. Meanwhile,  SemStamp and Adatpive changes sentences more remarkably, resulting in a significant deviation from the original output. For instance, in Semstamp, the phrase ``are suing the college" is entirely replaced by ``alleging they were misled about their ability to become certified", modifying the sentence's meaning. While Adatpive introduced a new phrase of ``have filed a suit claiming that Newbridege made “materially incorrect” claims'', causing the output to deviate from the original text. In addition, we observe a tiny portion of green highlights in these two WMs, indicating  substantial dissimilarities in the outputs. Also, the perplexity of the  SemStamp and Adaptive WM are considerably higher at $10.61$ and $17.62$ respectively, reflecting the degradation in model utility and semantic consistency.

\begin{table*}[t!]
\centering
\caption{Selected prompt and outputs from the LLaMA-2 model without and with WMs. Green text indicates semantic similarities of WM text and the original output without WMs.}
\begin{tabular}{|p{0.11\linewidth}|p{0.77\linewidth}|p{0.045\linewidth}|}
\hline
\textbf{} & \centering \textbf{Output text} & \textbf{PPL} \tabularnewline \hline
\centering \textbf{Prompt} & 
\raggedright ...medical laboratory technician associate program. FILING SUIT: Attorney Scott Schutzman discusses a class-action lawsuit against Newbridge College with two of his clients. A group of students... & 
\centering  \tabularnewline \hline
\centering  \multirow{3
}{*}{\makecell{\textbf{No}\\ \textbf{WM}}}  & 
\raggedright \hl{are suing the college}, claiming its \hl{medical laboratory technician program} is not accredited and does not qualify them to \hl{work in the field}. \hl{SANTA ANA} – \hl{A group of students} have \hl{filed a class-action lawsuit} \hl{against Newbridge College}, claiming the \hl{college}'s \hl{medical laboratory technician program} is [...continues] & 
\centering \multirow{3}{*}{\makecell{2.32}}  \tabularnewline \hline
\centering \multirow{4
}{*}{\makecell{\textbf{KGW}}} & 
\raggedright \hl{are suing the college} for allegedly misrepresenting the quality of its \hl{medical laboratory technician program}. \hl{Newbridge College} faces \hl{lawsuit} over alleged false advertising of \hl{medical lab technician program} \hl{SANTA ANA} – \hl{A group of students} have \hl{filed a class-action lawsuit} \hl{against Newbridge College}, alleging the \hl{college} misrepresented [...continues] & 
\centering \multirow{4}{*}{\makecell{2.75}}  \tabularnewline \hline
\centering \multirow{3
}{*}{\makecell{\textbf{EXP}}} & 
\raggedright sued \hl{the college} claims the school's program does not \hl{qualify} them to be \hl{medical lab technicians}. STUDENTS: ATTORNEY SCOTT SCHUTZMAN: Two of the \hl{Newbridge College} students who \hl{filed a class-action lawsuit} against the school are pictured with their attorney, Scott Schutzman. [...continues] & 
\centering \multirow{3}{*}{\makecell{4.43}}  \tabularnewline \hline
\centering
\multirow{4}{*}{\makecell{\textbf{SIR}}} & 
\raggedright sued \hl{the college} for false advertising after they say the college's \hl{medical laboratory technologist program} does not prepare them for the job market. CENTER: \hl{Newbridge College}'s Santa Ana campus is located at 2840 East 17Th Street. Students who attended the college have \hl{filed a class- action lawsuit} against the college alleging false advertising.\hl{SANTA ANA} - Students who enrolled in \hl{Newbridge College} [...continues] & 
\centering \multirow{4}{*}{\makecell{5.11}}  \tabularnewline \hline
\centering \multirow{4
}{*}{\makecell{\textbf{SemStamp}}} & 
\raggedright alleging they were misled about their ability to become certified as \hl{medial labs techs} through the \hl{school's programs} and \hl{have filed suit} for damages. (File Photo) Newport Beach lawyer files complaint on behalf of former NBCC student who claims she was defrauded out of thousands of dollars due to false promises made by institution regarding job placement rate  [...continues] & 
\centering \multirow{4}{*}{\makecell{10.61}}  \tabularnewline \hline
\centering \multirow{4
}{*}{\makecell{\textbf{Unigram}}} & 
\raggedright \hl{sued the college} for allegedly making false claims about its \hl{medical lab technician program}.
STUDENTS: Attorney Scott Schutzman represents two \hl{students}, Alexis Mesa and John Doe, who \hl{sued Newbridge College} for allegedly making false claims about its \hl{medical lab technician program}.  [...continues] & 
\centering \multirow{4}{*}{\makecell{4.11}}  \tabularnewline \hline
\centering \multirow{4
}{*}{\makecell{\textbf{Adaptive}}} & 
\raggedright have \hl{filed a suit} claiming that Newbridege made ``materially incorrect'' claims about it programs, such as saying it offered a \hl{medical lab technician program} that \hl{qualified} them to be medical lab techs. Atty: New Bridge falsely advertised program. By Joel Kimber | Orange County Register. A Newport Beach lawyer has filed a proposed \hl{class-action lawsuit} on behalf of students who say they were duped by \hl{Newbridge College}  [...continues] & 
\centering \multirow{4}{*}{\makecell{17.62}}  \tabularnewline \hline
\centering \multirow{4
}{*}{\makecell{\textbf{UPV}}} & 
\raggedright have \hl{sued the college} for fraud and false advertising, \hl{claiming the college's medical laboratory technology program} does not prepare them for \hl{jobs in the field}. SCHUTZMAN: Attorney Scott Schutzman represents the students who have sued \hl{Newbridge College}, accusing the college of fraud and false advertising. [...continues] & 
\centering \multirow{4}{*}{\makecell{4.27}}  \tabularnewline \hline
\end{tabular}
\label{table:viz_LLM2_text}
\end{table*}
\setlength{\textfloatsep}{10pt}

\newpage

\subsection{Remarks}
Through extensive experiments, we observe several key insights as follows. 
\begin{itemize}
    \item WMs significantly enhance the classification of LLM outputs, supporting the generation of unique and differentiated outputs. This enables clouds to identify their outputs and protect proprietary content accurately. By using WMs, clouds can strengthen IP protection and reduce unauthorized use, which is vital in competitive industries.
    \item  WMs typically have a moderate impact on model utility, subtly increasing perplexity in text generation tasks while reducing accuracy in MMLU downstream tasks. The semantic meaning of outputs is mostly preserved after WM, with significant changes occurring mostly in SemStamp and Adaptive. However, it is worth noting that LLM outputs remain beneficial 
    even after being watermarked.
    \item Attacks adversely affect WMs by degrading the WMs' detection function, compromising WMs'   effectiveness.  However, it comes with a considerable cost to model utility. 
    Further investigation is necessary to understand the trade-offs involved and to develop strategies that enhance WM resilience while maintaining model utility. 
\end{itemize}

\section{Challenges and Future Directions}   
While WMs demonstrate promising effectiveness across various LLMs, several challenges and barriers remain, necessitating further investigation before deploying them in the real-world applications. 

\subsection{Different Sensitivities among Output Tokens}
Our experiments reveal that tokens exhibit varying levels of sensitivity, especially after being watermarked.  As shown in Table \ref{table:sensitivity}, numerical tokens are changed significantly compared with other tokens. For instance, the phrase ``Nasdaq dropped 0.7 percent" in the original output is altered to ``Nasdaq lost 0.6 percent" in the EXP, SIR and Adaptive, to ``Nasdaq lost 1.2 percent" in Unigram, and to ``Nasdaq lost 1.4 percent" in UPV. Similarly, the original output states that the U.S. economy grew at a ``2.5 percent annual rate in the second quarter", which is shifted to ``1 percent" or ``3.5 percent" with WMs. These  unintended modifications to numerical tokens result in changing the semantic meaning of the output. This change poses significant risks in domains where precise values are important, including financial reporting, scientific data analysis, weather forecasting, and legal documentation. As a result, there is a pressing need to address these challenges to ensure that WMs do not compromise the integrity of essential information.

\begin{table*}[t!]
\centering
\caption{This example demonstrates the sensitivity of numerical tokens in LLaMA-2 outputs without and with different types of WMs. Yellow highlights represent key numerical information in the output without WMs, while green highlights indicate similarities, and red ones are for differences compared to no WM setting.}
\begin{tabular}{|p{0.11\linewidth}|p{0.85\linewidth}|}
\hline
\textbf{} & \centering \textbf{Content} \tabularnewline \hline
\centering \textbf{Prompt} & 
\raggedright ...For the week, the Dow fell 0.8 percent, the S\&P 500 slipped 0.9 percent and the ... \tabularnewline \hline
\centering  \multirow{3
}{*}{\makecell{\textbf{No}\\ \textbf{WM}}}  & 
\raggedright Nasdaq dropped \hlyellow{0.7 percent}. The report from the Commerce Department showed that the U.S. economy grew at a slower-than-expected \hlyellow{2.5 percent} annual rate in the \hlyellow{second quarter}, down from a \hlyellow{3.4 percent} pace in the \hlyellow{first quarter}. [...continues] \tabularnewline \hline
\centering \multirow{4
}{*}{\makecell{\textbf{KGW}}} & 
\raggedright Nasdaq lost \hl{0.7 percent}. The economic report from the Commerce Department showed that the U.S. economy grew at a slower-than-expected \hlred{3.5 percent} annual rate in the \hlred{first quarter}, down from a previously estimated \hlred{4.4 percent pace}. [...continues]  
\tabularnewline \hline
\centering \multirow{3
}{*}{\makecell{\textbf{EXP}}} & 
\raggedright Nasdaq lost \hlred{0.6 percent}. The U.S. economy grew at a slower-than-expected \hlred{1.9 percent} annual rate in the \hlred{first quarter}, the Commerce Department reported. Economists had been expecting a rate of \hlred{2.2 percent}. [...continues] 
\tabularnewline \hline
\centering
\multirow{3}{*}{\makecell{\textbf{SIR}}} & 
\raggedright Nasdaq lost \hlred{0.6 percent}. The Commerce Department said Friday that the U.S economy grew at a \hl{2.5 percent} annual rate in the \hlred{first quarter}, slower than the \hlred{4.4 percent} rate in the previous quarter [...continues] 
\tabularnewline \hline
\centering \multirow{3}{*}{\makecell{\textbf{SemStamp}}} & 
\raggedright DAQ declined by less than \hlred{half} that amount as investors grew cautious amid signs global demand is weakening despite low interest rates worldwide. [...continues] 
\tabularnewline \hline
\centering \multirow{4
}{*}{\makecell{\textbf{Unigram}}} & 
\raggedright Nasdaq dropped \hlred{1.2 percent}. The disappointing economic data came from the Commerce Department (search), which said gross domestic product (GDP) grew at a seasonally adjusted annual rate of \hlred{3.3 percent} in the \hlred{first quarter}, slower than the \hlred{4.8 percent} pace in the previous quarter. [...continues]  
\tabularnewline \hline
\centering \multirow{4
}{*}{\makecell{\textbf{Adaptive}}} & 
\raggedright NasdaQ dropped \hlred{0.6 percent}. The U.S economy expanded at a \hlred{1 percent} pace in the \hlred{first quarter}, according to a revised estimate from the government, which is below expectations of \hlred{2 percent to 2.5 percent}. It also reflected a downward revision to growth in the previous quarter.[...continues] \tabularnewline \hline
\centering \multirow{4
}{*}{\makecell{\textbf{UPV}}} & 
\raggedright Nasdaq dropped \hlred{1.4 percent}. The report on U.S. gross domestic product (GDP) released by the Commerce Department showed the economy grew at an annual rate of \hl{2.5 percent} in the \hl{second quarter}, below the \hlred{3.5 percent} pace of the \hl{first quarter} [...continues] \tabularnewline \hline
\end{tabular}
\label{table:sensitivity}
\end{table*}
\setlength{\textfloatsep}{10pt}

\subsection{Robustness to Removal Attacks}
WMs, regardless of their category, generally face challenges related to  robustness. 
 Given the complex manipulation behaviors of users and adversaries  in the real-world, it is difficult to create a WM that can resist to modifications such as paraphrasing, lexical changes, or reordering. As also claimed  in \cite{zhang2023watermarks}, current strong WMs  fail to offer the desired level of security, especially against sophisticated attacks.

\subsection{Text Quality and Semantic Drift}
Embedding WMs in LLMs, whether by modifying logits,  sampling process, or updating model weights, disrupts the natural flow of text generation. In Table \ref{table:PPL_scores}, even KGW, which has the smallest impact on model utility, still alters the original perplexity by $9-25\%$.  This disruption adversely affects  the naturalness and fluency of the watermarked output, leading to a overall reduction in model utility and text quality, compared with those of original outputs. In sensitive contexts, semantic drift can further compromise the credibility and usefulness of LLM outputs. Given the inevitability of the issue, it is essential to develop methods for quantifying the impact and managing it in acceptable thresholds. 

\subsection{Scalability and Practicality}
The scalability and practicality of WMs for LLMs pose another significant challenge. WMs often require modifications to the model's generation process, which might cause higher processing time and resource consumption. In addition, some WMs require additional training to function properly. As LLMs are used on a broad scale, particularly in real-time applications such as chatbots or content generating systems, the increased computational overhead may become prohibitive.
In addition, ensuring that WMs are robust across various domains, tasks, and languages complicates their practical deployment. Scaling these methods to handle large volumes of generated text while maintaining text quality and performance remains an open challenge, especially given the numerous use cases of LLMs across platforms. Future advancements should address these difficulties  to make WMs more realistic.

\subsection{WMs with Certified Robustness}
There has been no systematic work that formally guarantees the effectiveness of WMs for LLMs. Therefore, it is essential to develop WMs with certified robustness, where the magnitude of the WMs is explicitly measured in relation to the accuracy of generated outputs in both generation tasks and downstream tasks. This approach would ensure that WMs remain identifiable even in adversarial settings, while maintaining a defined threshold for output quality. By quantifying this relationship, we can create a system that balances robustness and model utility, offering a more reliable and scalable method for watermarking LLMs.

\subsection{Explore WMs and Alignment in LLMs}
Alignment refers to ensuring that LLMs generate outputs consistent with standard ethical and operational guidelines \cite{shen2023large}. However,  WMs can interfere with this alignment by altering the generation or model training process, potentially reducing the model's adherence to its pre-trained alignment. This issue becomes particularly critical in sensitive domains where the model's behaviors must align  with strict safety or ethical standards.
The impact of WMs on LLM alignment demands careful investigation, as disruptions in logits, sampling, or model weights could cause semantic drift or unintended biases in LLM outputs \cite{verma2025watermarking}. 

\section{Conclusion}
We presented an SoK for WMs in LLMs, with a particular focus on their potential for real-world deployment through a comprehensive evaluation. Our work encompasses a broader spectrum, including assessments of model utility in text generation and downstream tasks, semantic preservation of output under perplexity observation and qualitative assessment, robustness under various attack types, and  watermark sensitivity across LLMs, and consider other practical aspects.

To capture the intricate relationships among WMs, LLMs, and model utility, we introduced a novel cross-model IP classifier and examone the impacts of WMs on LLM output quality. 
 Our experiments demonstrate that WMs significantly enhance the identification of LLM outputs, strengthening IP protection and reducing unauthorized use. However, while WMs are promising for practical applications, challenges remain due to potential impacts on model utility, text quality, and vulnerabilities to WM removal attacks.  
Our discussion highlights the current state of research, identifies key limitations, and outlines future directions to improve the robustness, scalability, and practicality of WMs. We emphasize the need for further investigation to address these challenges and make watermarking techniques more viable for deployment across a range of LLM applications.

\newpage
\clearpage

\bibliographystyle{IEEEbib}
\bibliography{SoK}

\begin{figure}[t] 
\centering
\subfloat[Unigram]{\label{fig:subnets-aaa}\includegraphics[scale=0.305]{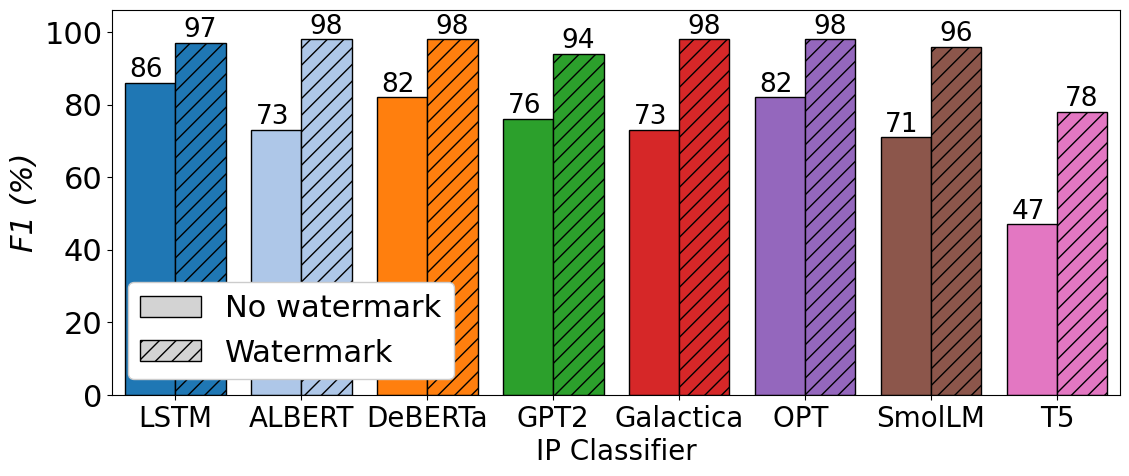}}\hfill
\subfloat[Adaptive]{\label{fig:subnets-bbb}\includegraphics[scale=0.305]{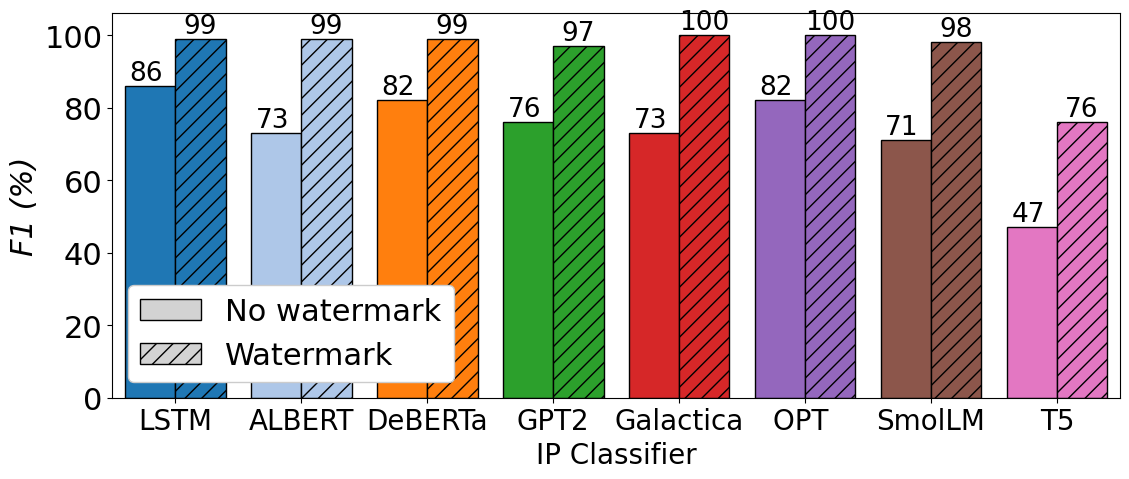}}\hfill
\subfloat[UPV]{\label{fig:subnets-ccc}\includegraphics[scale=0.305]{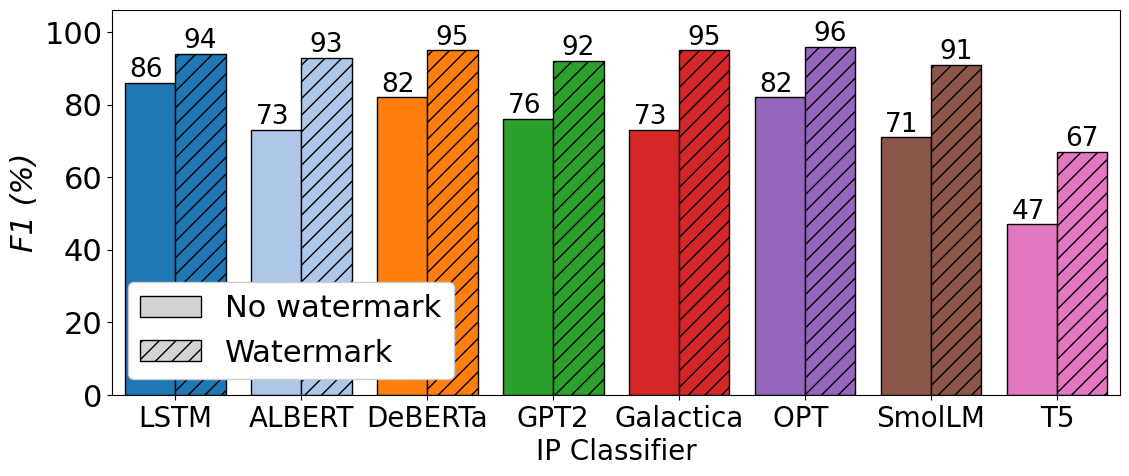}}
\caption{$F1$ scores as a function of different IP classifiers for different WMs, compared with No WM settings (Additional to results in Figure \ref{fig:IPChecker}).}
\label{fig:IPChecker_addition}  
\end{figure}

\appendix
\noindent \textbf{Impacts of other WMs on IP Classifiers.} 
As we applied multiple WMs to reassure our assessment, the cross-model IP classifier results of Unigram, Adaptive and UPV WMs are shown in Fig.~\ref{fig:IPChecker_addition}. Similar insights could be drawn from the figure, reaffirming our observation of the study.


\noindent \textbf{APIs for Deployment of Major LLM Providers. }  
To provide context on the current deployment status of LLMs, we present Table \ref{table:llmAPI2}, which compares various real-world APIs, outlining their applications, advantages, and limitations. Leading providers like OpenAI, Microsoft, Meta, Hugging Face, Google, and Amazon offer a range of LLM capabilities, including NLP, sentiment analysis, code generation, and model fine-tuning. While each API offers strengths such as scalability, integration, and community support, they also come with challenges like high resource demands, usage-based pricing, and limited customization options.

\begin{table*}[t!]     
\centering
\caption{LLM Real-world APIs. }
\begin{tabular}{|p{0.115\linewidth}|p{0.08\linewidth}|p{0.05\linewidth}|p{0.19\linewidth}|p{0.21\linewidth}|p{0.2\linewidth}|}
\hline
\textbf{API Name} &  \textbf{Provider} & \textbf{Access Type} &  \textbf{Applications} & \textbf{Advantages} & \textbf{Constraints} \\ \hline
 \hline
ChatGPT and GPT family  \cite{OpenAI} & OpenAI & API &  NLP, text generation, questions and answers, chatbots & State-of-the-art LLMs, versatility, efficiency, personalization, cost-effective & Lack of real-Time knowledge, cost for high usage \\ \hline
Microsoft Azure Language \cite{MicrosoftAzureAI} & Microsoft Azure & API & Sentiment analysis, entity recognition, language understanding & Comprehensive APIs, scalability, integration with Azure ecosystem  & Pricing based on usage, limited offline capabilities  \\
\hline
LLaMA family \cite{HuggingFaceMetaLLaMA} & Meta & Open source & Text/code generation, text summarization, translation, recommendations, chatbots & High model utility, versatility, efficiency, fine-tuning support & Resource intensive, overfitting, high maintenance and updates\\ \hline
Hugging Face transformers \cite{HuggingFaceTransformers} & Hugging Face & Open source &  NLP, model fine-tuning, text generation, research and development & large model repository, extensive community support,  interoperability, versatility & No official support, resource intensive \\
\hline
Google Cloud AI-Language (Gemini, Bard, PaLM, etc.) \cite{GoogleCloudNLP} & Google & API  & Sentiment analysis, entity recognition, translation, dialogue, text/code/image generation, comprehension & Google's robust and substantial infrastructure, easy integration with Google services, proficiency & Pricing based on usage, quality of generated content, inconsistent responses \\
\hline
Amazon Comprehend \cite{satyanarayana2020sentimental}  & Amazon AWS & API &  Entity recognition, sentiment analysis,  classification,  topic modeling &  Integration with other AWS services, scalability, reliability & Pricing based on usage, limited customization \\
\hline
IBM Watson NLU \cite{IBMNLP} & IBM Watson & API & Sentiment and emotion analysis, keyword extraction & Comprehensive text analysis, customization, scalability, rich documentation and support,  Integration with other IBM services & Cost for high usage, overhead for simple tasks\\
\hline
 Fairseq \cite{satyanarayana2020sentimental} & Meta & Open source & Translation, research and development, language modeling &  Versatility, research-friendly, integration with PyTorch & Limited user community, less user-friendly for beginners, maintenance \\
 \hline Claude \cite{AnthropicClaude} & Anthropic & API &  Summarization, search, collaborative writing,  coding comprehension  & Safety and alignment focus, human-like interaction, multi-modality  & Costly, limited customization \\
 \hline
 Cohere \cite{Cohere} & Cohere Inc. & API &   Chatbots, knowledge assistants, scalability & User-friendly, seamless customization, focus on enterprise needs, privacy & Costly, small community\\
  \hline
\end{tabular}
\label{table:llmAPI2}
\end{table*}

\begin{table*}[t]
\centering
\caption{Details on WM Schemes used in Experiments. }
\begin{tabular}{|p{0.076\linewidth}|p{0.46\linewidth}|p{0.41\linewidth}|}
\hline
\textbf{\small{WM}} & \textbf{\small{Details}} & \textbf{\small{Mechanism Visualization}} \\
\hline
\centering 
\multirow{2}{*}{\makecell{\textbf{KGW} \cite{kirchenbauer2023watermark}}}
 & \begin{tabular}{l}
  Divides the vocabulary into green and red tokens, then  \\
 modifies  logits by adding a fixed WM  to  green  \\
 tokens' logits to favor the generation of green tokens. 
  \end{tabular}  
 & 
 \begin{minipage}{.3\textwidth}
      \includegraphics[scale=0.022]{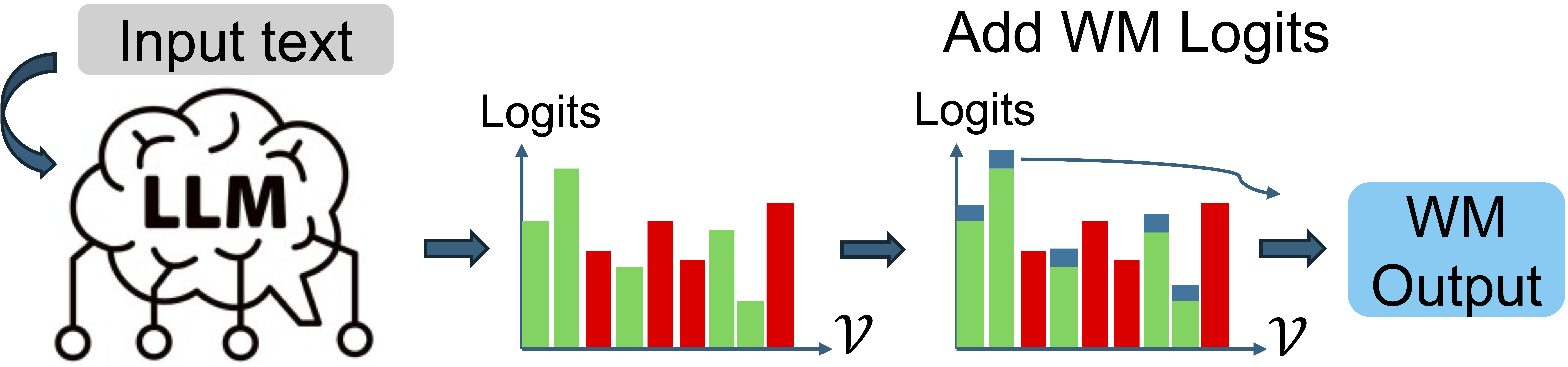}
    \end{minipage} \\
\hline
\centering \multirow{1.5
}{*}{\makecell{\textbf{SIR} \cite{liu2024semanticinvariantrobustwatermark}}}  &  \begin{tabular}{l}
   Determines WM logits by converting  semantics  \\
    of all preceding tokens using an auxiliary embedding\\
       model,  then combines them with the  original logits.
  \end{tabular}
  & 
 \begin{minipage}{.3\textwidth}
      \includegraphics[scale=0.022]{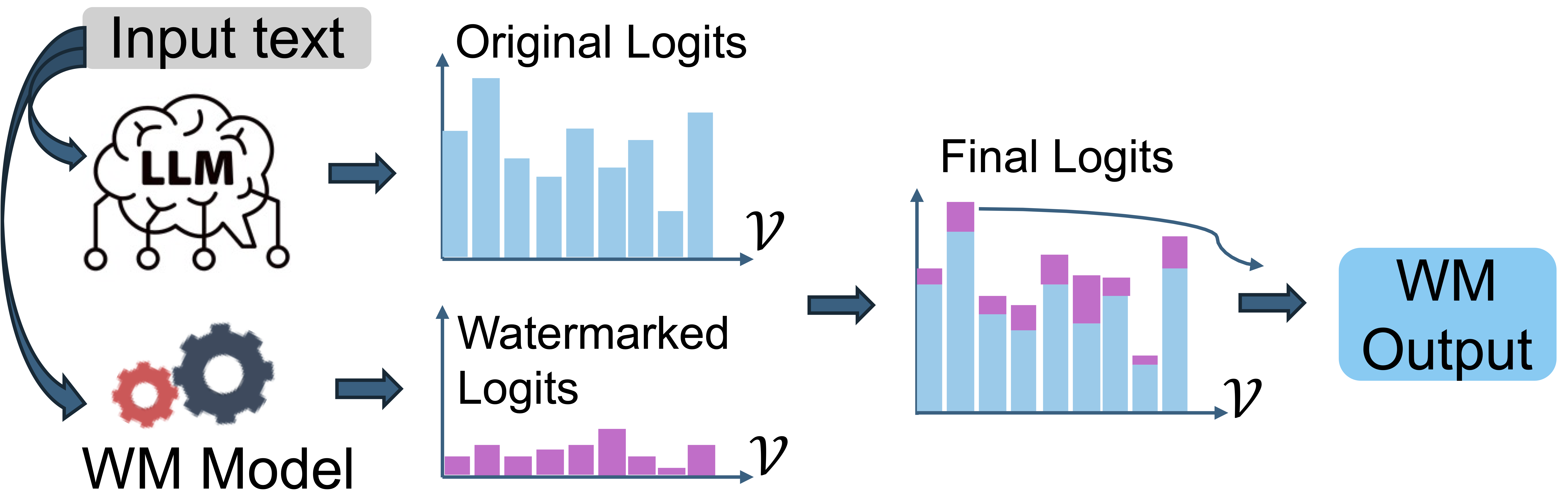}
    \end{minipage} \\
\hline
\centering \multirow{1.5
}{*}{\makecell{\textbf{EXP} \cite{kuditipudi2023robust}}} & \begin{tabular}{l}
   Intervenes the sampling process of each token by  \\
   mapping  it with a pseudo-random number sequence, in   \\
   which every generated token corresponds with a key.
  \end{tabular}
& \begin{minipage}{.3\textwidth}
      \includegraphics[scale=0.022]{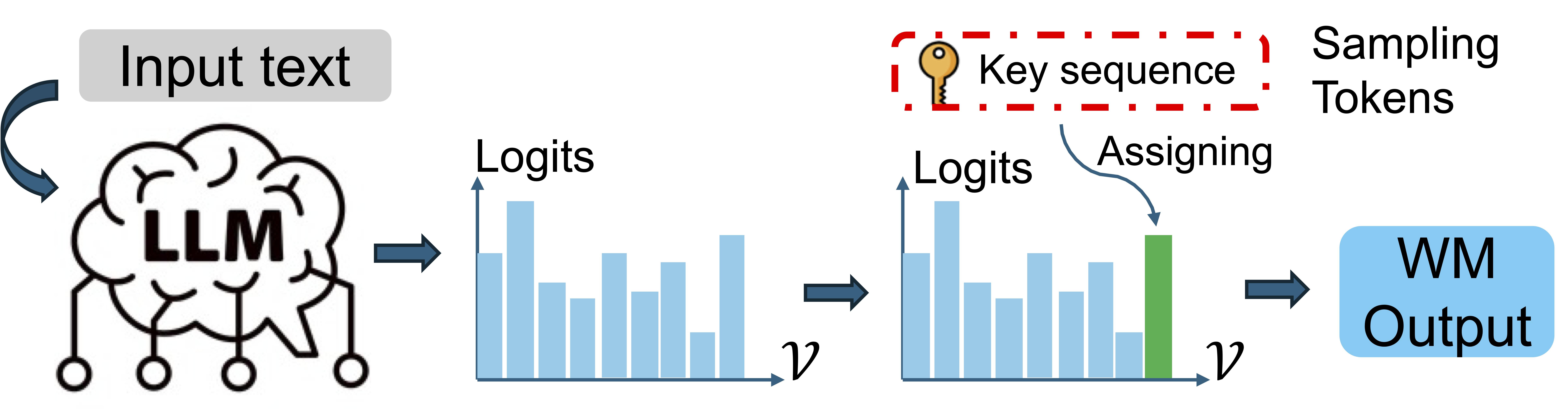}
    \end{minipage} \\
\hline
\centering \multirow{1.5
}{*}{\makecell{\textbf{\small{SemStamp}}\\ \cite{hou2023semstamp}}}   & \begin{tabular}{l}
   Generates new sentences by mapping candidates into \\ 
   embeddings, partitions  semantic space into valid \& \\
  blocked regions, and chooses  sentences in  valid regions.
  \end{tabular}
 & 
\begin{minipage}{.3\textwidth}
      \includegraphics[scale=0.022]{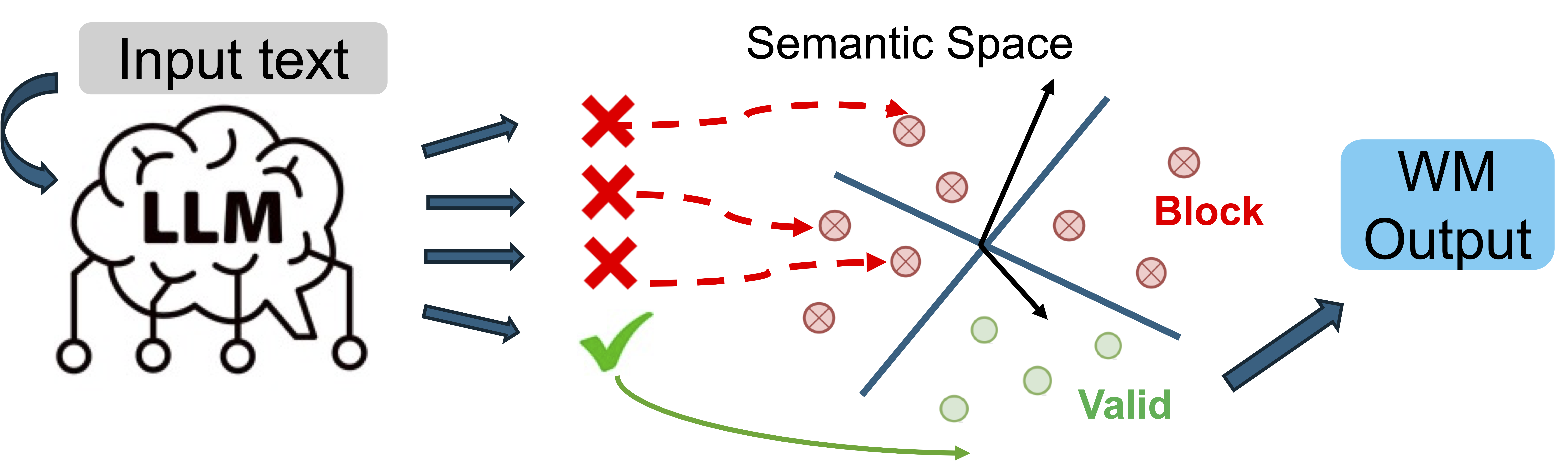}
    \end{minipage} \\
\hline
\centering \multirow{1.5
}{*}{\makecell{\textbf{\small{Unigram}}\\ \cite{zhao2024provable}}}   & \begin{tabular}{l}
Modifies the logits to embed the WM by using \\a 
fixed grouping strategy, same green and red lists \\ 
applied consistently for entire vocabulary.
  \end{tabular}
 & 
\begin{minipage}{.3\textwidth}
      \includegraphics[scale=0.0215]{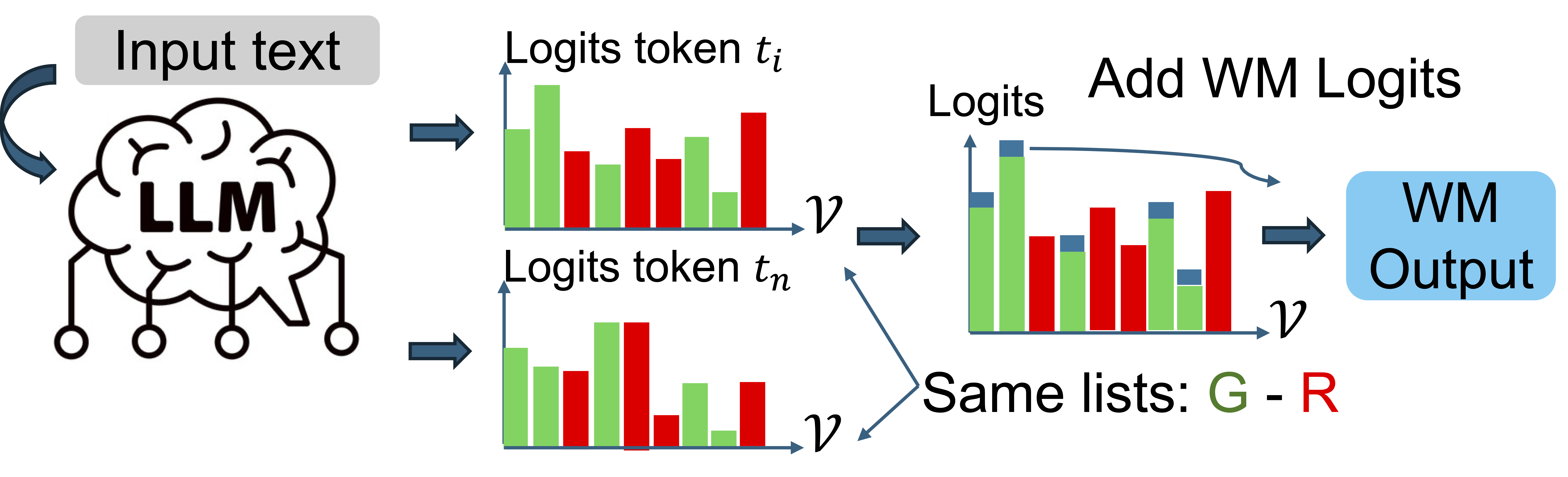}
    \end{minipage} \\
\hline
\centering \multirow{1.5
}{*}{\makecell{\textbf{\small{Adaptive}}\\ \cite{liu2024adaptive}}}   & \begin{tabular}{l}
Adaptively scales up logits of high-entropy tokens \\
based on preceding tokens' semantics embedding.
  \end{tabular}
 & 
\begin{minipage}{.3\textwidth}
      \includegraphics[scale=0.020]{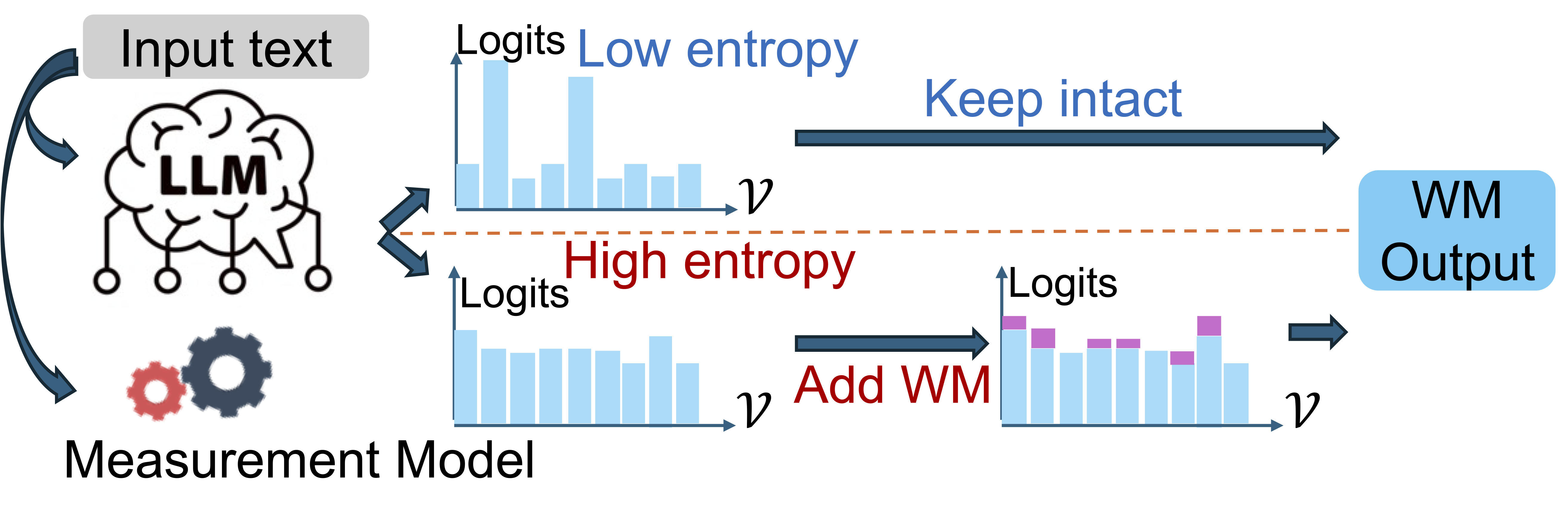}
    \end{minipage} \\
\hline
\centering \multirow{1.5
}{*}{\makecell{\textbf{UPV} \cite{liu2023unforgeable} }}   & \begin{tabular}{l}
Employs two networks for WM generation (modifying \\
logits) and detection (evaluating entire text without \\a generation key).
  \end{tabular}
 & 
\begin{minipage}{.3\textwidth}
      \includegraphics[scale=0.022]{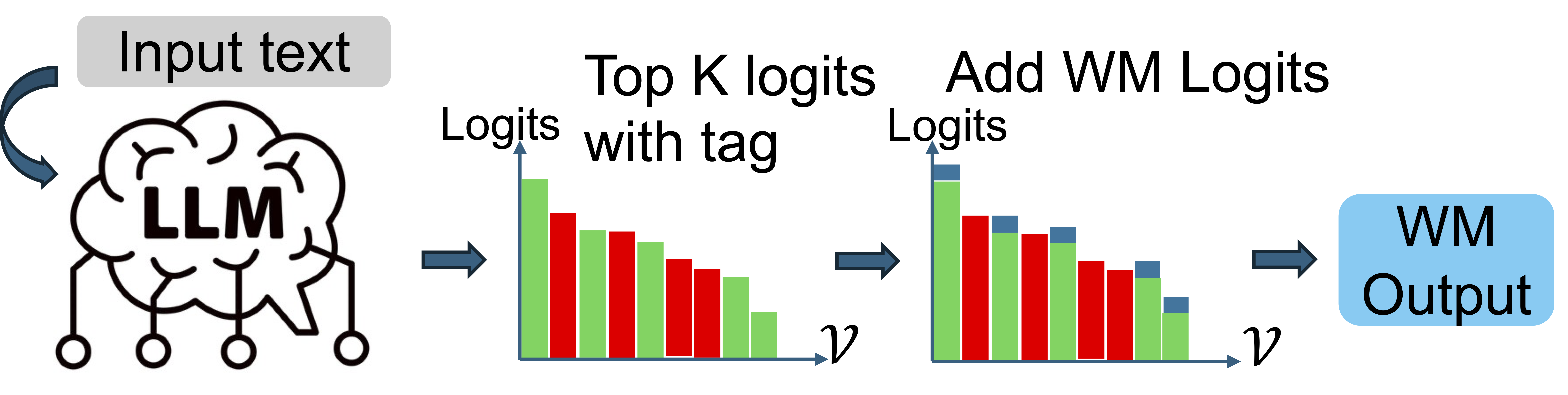}
    \end{minipage} \\
\hline
\end{tabular}
\label{table:7WMs}
\end{table*}


\end{document}